\documentclass[twocolumn,english]{revtex4-2}
\usepackage{lmodern}
\usepackage{lmodern}

\usepackage[T1]{fontenc}
\usepackage[latin9]{inputenc}
\usepackage{babel}
\usepackage{amsmath}
\usepackage{amssymb}
\usepackage{graphicx}
\usepackage[unicode=true,pdfusetitle,
 bookmarks=true,bookmarksnumbered=false,bookmarksopen=false,
 breaklinks=false,pdfborder={0 0 1},backref=false,colorlinks=false]
 {hyperref}

\makeatletter

\providecommand{\tabularnewline}{\\}
\newcommand{\lyxdot}{.}

\usepackage{babel}

\makeatother

\begin{document}
\title{Piston driven shock waves in non-homogeneous planar media}
\author{Menahem Krief}
\email{menahem.krief@mail.huji.ac.il}

\address{Racah Institute of Physics, The Hebrew University, 9190401 Jerusalem,
Israel}
\begin{abstract}
In this work, we analyze in detail the problem of piston driven shock
waves in planar media. Similarity solutions to the compressible hydrodynamics
equations are developed, for a strong shock wave, generated by a time
dependent pressure piston, propagating in a non-homogeneous planar
medium consisting of an ideal gas. Power law temporal and spatial
dependency is assumed for the piston pressure and initial medium density,
respectively. The similarity solutions are written in both Lagrangian
and Eulerian coordinates. It is shown that the solutions take various
qualitatively different forms according to the value of the pressure
and density exponents. We show that there exist different families
of solutions for which the shock propagates at constant speed, accelerates
or slows down. Similarly, we show that there exist different types
of solutions for which the density near the piston is either finite,
vanishes or diverges. Finally, we perform a comprehensive comparison
between the planar shock solutions and Lagrangian hydrodynamic simulations,
by setting proper initial and boundary conditions. A very good agreement
is reached, which demonstrates the usefulness of the analytic solutions
as a code verification test problem.
\end{abstract}
\maketitle

\section{Introduction}

Shock waves are an important phenomena in many astrophysical and laboratory
high energy density plasmas \cite{lindl2004physics,robey2001experimental,bailey2015higher,falize2011similarity,hurricane2014fuel,abu2022lawson}.
As a result, analytic solutions for the compressible hydrodynamics
equations play a key role in the analysis and design of high energy
density experiments \cite{sigel1988x,keiter2008radiation,back2000diffusive,lindl1995development,heizler2021radiation,cohen2020key}
and in the process of verification and validation of computer simulations
\cite{calder2002validating,ramsey2018converging,ramsey2019piston,ruby2019boundary,giron2021solutions,modelevsky2021revisiting,krief2021analytic,robert2021determination,chauhan2021piston,calvo2022stability,ganapa2021blast}. 

In Ref. \cite{shussman2015full}, self-similar solutions for ablation
driven shock waves \cite{pakula1985self,kaiser1989x,hammer2003consistent,saillard2010principles,garnier2006self,clarisse2018hydrodynamic,malka2022supersonic}
in homogeneous media, are presented. The ablation pressure serves
as a pressure piston boundary condition for the shock wave, with a
given temporal power-law dependency of the form $p_{0}t^{\tau}$.
The separate treatment of the ablation region and the shock region
enables the use of a binary-equation of state which is essential for
an accurate modeling of some real solid materials experiments \cite{heizler2016self}.
For example, this solution enables an analytic theoretical modeling
of subsonic radiative heat flow, used for evaluating the drive temperature
in hohlraum experiments via a measurement of the shock wave velocity
in aluminum wedges \cite{heizler2021radiation}. 

The ablation driven shock solutions \cite{shussman2015full}, assume
an ideal gas equation of state (EOS) and a medium with an initial
homogeneous density. In this work we extend this problem and develop
similarity solutions to the compressible hydrodynamics equations for
a non-homogeneous medium with an initial density profile which is
a spatial power law of the form $\rho_{0}x^{-\omega}$, as described
in Fig. \ref{fig:problem_desc}. Power law density profiles are widely
used in many fields. For example, in astrophysical modeling, as the
interior and atmospheres of stars \cite{meszaros2002theories,tan2001trans,sapir2011non,katz2012non,sapir2013non,remorov2022propagation}
and galaxies \cite{evans1994power,koopmans2009structure,schneider2013mass}. 

The generalized shock solutions are presented and analyzed in detail.
Different qualitative behavior of the propagation and structure of
the piston driven shock wave is shown to exist for different ranges
of the temporal exponent $\tau$ and spatial exponent $\omega$. It
is shown that the shock speed could be constant, accelerate and decelerate
and that the density near the piston can be finite, vanish or diverge.
The solutions are studied in both Lagrangian mass coordinates and
in real space. We show that the transformation between these two representations
is given by a closed formula, which demonstrates the behavior of the
shocked fluid in the $x-t$ plane. We study the temporal dependence
of the energy supplied to the fluid by the pressure load, and how
it is divided into kinetic and internal fluid energy.

Finally, we perform a comprehensive comparison between the semi-analytical
planar shock solutions to numerical hydrodynamic simulations. As shown,
a very good agreement is reached, which highlights the use of the
new generalized solutions for the purpose of verification and validation
of numerical hydrodynamics simulation codes. In addition, the solutions
presented in this work can be used directly to generalize the ablation
driven shock wave solutions \cite{shussman2015full,heizler2021radiation},
to the case of an initially non-homogeneous media. 

\section{Statement of the problem}

\begin{figure}[t]
\begin{centering}
\includegraphics[scale=0.7]{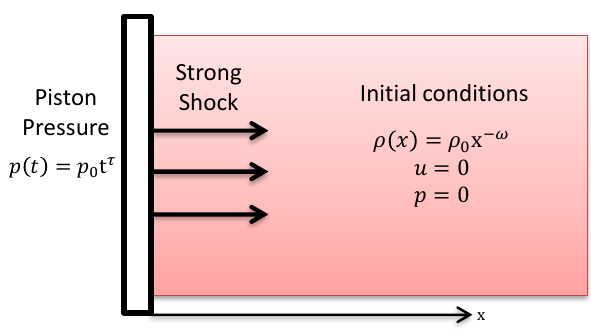}
\par\end{centering}
\caption{A schematic description of the hydrodynamics problem: a time dependent
pressure piston drives a strong shock wave, into a medium which is
initially at rest with a power law density profile.\label{fig:problem_desc}}
\end{figure}

In this section we will list the governing equations of motion and
boundary conditions for the hydrodynamic problem consisting of a piston
which exerts a time dependent external pressure on a planar medium
which is initially at rest, with an initial spatial power law density
profile. The time dependent pressure is assumed to be a temporal power
law.  This problem is described schematically in Fig. \ref{fig:problem_desc}. 

Since the system boundary changes over time, it is customary to work
in Lagrangian mass coordinates:
\begin{equation}
m\left(x,t\right)=\int_{0}^{x}\rho\left(x',t=0\right)dx',
\end{equation}
where $\rho\left(x,t\right)$ is the planar mass density. The hydrodynamic
equations in planar symmetry and Lagrangian mass coordinates are given
by (conservation of mass, momentum and energy, respectively):
\begin{equation}
\frac{\partial v}{\partial t}-\frac{\partial u}{\partial m}=0,\label{eq:_mass_hydro}
\end{equation}
\begin{align}
\frac{\partial u}{\partial t} & +\frac{\partial p}{\partial m}=0,\label{eq:p_hydro}
\end{align}

\begin{equation}
\frac{\partial e}{\partial t}+p\frac{\partial v}{\partial t}=0,\label{eq:energy_hydro}
\end{equation}
where $v\left(m,t\right)=1/\rho\left(m,t\right)$ is the specific
volume, $u\left(m,t\right)$ is the fluid velocity, $p\left(m,t\right)$
is the pressure and $e\left(m,t\right)$ is the specific internal
energy (energy per unit mass). In this work we will assume a polytropic
ideal gas equation of state with an adiabatic index $\gamma=r+1$
so that:
\begin{equation}
pv=re.
\end{equation}
The boundary condition consists of a pressure piston with a power
law time dependence in the form:
\begin{equation}
p\left(m=0,t\right)=p_{0}t^{\tau}\label{eq:p_bc}
\end{equation}
The initial conditions are of a medium at rest:
\begin{equation}
u\left(m,t=0\right)=0,\ p\left(m,t=0\right)=0,\label{eq:init_conds_p_u}
\end{equation}
with a density profile which is a spatial power law of the form:
\begin{equation}
\rho\left(x,t=0\right)=\rho_{0}x^{-\omega},\label{eq:rho_init}
\end{equation}
The resulting initial specific volume is:
\begin{equation}
v\left(x,t=0\right)=v_{0}x^{\omega},
\end{equation}
where $v_{0}=1/\rho_{0}$. In order to write the initial specific
volume profile in terms of Lagrangian mass coordinates, we note that:
\begin{equation}
m\left(x,t=0\right)=\int_{0}^{x}\rho\left(x',t=0\right)dx'=\frac{1}{v_{0}}\frac{x^{1-\omega}}{1-\omega},\label{eq:minit}
\end{equation}
where we have used the fact that the total mass is finite, which leads
to the constraint:
\begin{equation}
\text{\ensuremath{\omega<1}}.
\end{equation}
Eq. \ref{eq:minit} gives the initial Eulerian to Lagrangian relation:
\begin{equation}
x\left(m,t=0\right)=\left[v_{0}\left(1-\omega\right)m\right]^{\frac{1}{1-\omega}},\label{eq:position_mnass_initial}
\end{equation}
and the initial specific volume in Lagrangian coordinates:
\begin{equation}
v\left(m,t=0\right)=v_{0}^{\frac{1}{1-\omega}}\left[\left(1-\omega\right)m\right]^{\frac{\omega}{1-\omega}}.\label{eq:init_conds_v_lag}
\end{equation}

\section{Self-Similar representation\label{sec:ss_rep}}

\begin{figure}[t]
\begin{centering}
\includegraphics[scale=0.6]{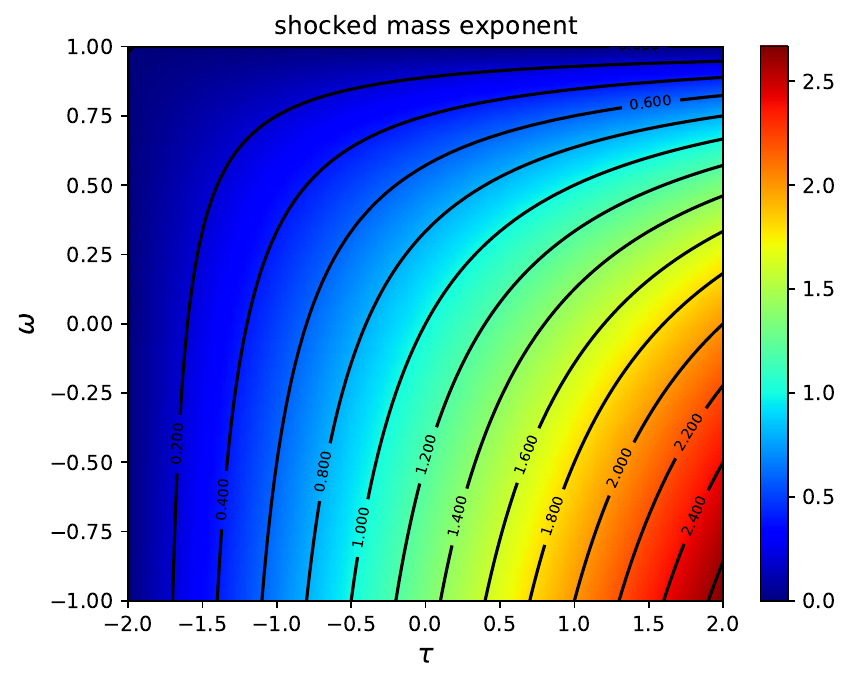}
\par\end{centering}
\caption{A color plot for the mass similarity exponent (the temporal power
of the shocked mass, see equation \ref{eq:shocked_mass}), as a function
of $\tau,\omega$.\label{fig:mass_pow}}
\end{figure}

The hydrodynamic problem defined in the previous section is defined
in terms of $N=7$ dimensional quantities: two independent variables
$m,t$, three dependent hydrodynamic profiles $p\left(m,t\right),u\left(m,t\right),v\left(m,t\right)$
and two dimensional parameters $v_{0},p_{0}$. The dimensions of these
quantities are listed in table \ref{tab:The-dimensional-parameters},
in terms of $K=3$ units: mass, length and time. Therefore, from the
central theorem of dimensional analysis (a.k.a. the Pi theorem) \cite{buckingham1914physically,zeldovich1967physics,barenblatt1996scaling},
the problem can be fully characterized in terms of $N-K=4$ dimensionless
variables, which are given in terms of power laws of the dimensional
quantities. Therefore, it is customary to use an independent dimensionless
similarity variable $\xi$ and three dependent dimensionless similarity
profiles $V\left(\xi\right),U\left(\xi\right),P\left(\xi\right)$
for the specific volume, velocity and pressure fields, respectively.
Writing these similarity variables in terms of power laws of the dimensional
quantities, yields the following self-similar representation:
\begin{equation}
\xi=mv_{0}^{\frac{1}{2-\omega}}p_{0}^{\frac{\omega-1}{2-\omega}}t^{\frac{\left(2+\tau\right)\left(\omega-1\right)}{2-\omega}},\label{eq:xsi_def}
\end{equation}
\begin{equation}
v\left(m,t\right)=V\left(\xi\right)v_{0}^{\frac{2}{2-\omega}}p_{0}^{\frac{\omega}{2-\omega}}t^{\frac{\left(2+\tau\right)\omega}{2-\omega}},\label{eq:v_ss}
\end{equation}

\begin{equation}
u\left(m,t\right)=U\left(\xi\right)v_{0}^{\frac{1}{2-\omega}}p_{0}^{\frac{1}{2-\omega}}t^{\frac{\omega+\tau}{2-\omega}},\label{eq:u_ss}
\end{equation}
\begin{equation}
p\left(m,t\right)=P\left(\xi\right)p_{0}t^{\tau}.\label{eq:p_ss}
\end{equation}
We note that the piston boundary condition \ref{eq:p_bc} is written
in terms of the similarity pressure profile as:

\begin{equation}
P\left(0\right)=1.\label{eq:p_bc_ss}
\end{equation}
By substitution of this self-similar representation in the hydrodynamics
equations \ref{eq:_mass_hydro}-\ref{eq:energy_hydro} and using the
relations $\frac{\partial\xi}{\partial m}=\frac{\xi}{m}$ and $\frac{\partial\xi}{\partial t}=\frac{\left(2+\tau\right)\left(\omega-1\right)}{2-\omega}\frac{\xi}{t}$,
all dimensional quantities factor out, and a set of nonlinear ordinary
differential equations for the similarity profiles is obtained:
\begin{equation}
\left(\frac{2+\tau}{2-\omega}\right)\left(\left(\omega-1\right)\xi V'+\omega V\right)-U'=0,\label{eq:eq_ss_mass}
\end{equation}

\begin{align}
\frac{\left(2+\tau\right)\left(\omega-1\right)}{2-\omega}\xi U'+U\left(\frac{\omega+\tau}{2-\omega}\right) & +P'=0,
\end{align}

\begin{align}
 & \frac{\left(2+\tau\right)\left(\omega-1\right)}{2-\omega}\frac{\xi}{r}VP'+\frac{\tau}{r}PV\label{eq:eq_ss_e}\\
 & +\left(1+\frac{1}{r}\right)\left(\frac{2+\tau}{2-\omega}\right)P\left(\left(\omega-1\right)\xi V'+\omega V\right)=0.\nonumber 
\end{align}
It is seen that for the special case of a spatially constant initial
density ($\omega=0$), the equations are reduced to the form given
in equations 43a-43c in Ref. \cite{shussman2015full}.

\begin{table}
\centering{}%
\begin{tabular}{|c|c|c|c|c|c|c|}
\hline 
$p$ & $u$ & $v$ & $v_{0}$ & $p_{0}$ & $m$ & $t$\tabularnewline
\hline 
\hline 
$ML\mathcal{T}^{-2}$ & $L\mathcal{T}^{-1}$ & $M^{-1}L$ & $M^{-1}L^{1-\omega}$ & $ML\mathcal{T}^{-\tau-2}$ & $M$ & $\mathcal{T}$\tabularnewline
\hline 
\end{tabular}\caption{The dimensional quantities in the problem, and their dimensions in
terms of mass ($M$) length ($L$) and time ($\mathcal{T}$) units.\label{tab:The-dimensional-parameters}}
\end{table}

\section{Jump Conditions\label{sec:jump}}

Since the medium is initially stationary with zero pressure, a strong
shock is formed due to the operation of the pressure piston. The hydrodynamic
profiles in the unshocked region are given simply by the initial conditions
(zero pressure and velocity and a power law density profile, see equations
\ref{eq:init_conds_p_u},\ref{eq:init_conds_v_lag}). In the shocked
region, the hydrodynamic profiles are obtained from the solution of
the similarity equations \ref{eq:eq_ss_mass}-\ref{eq:eq_ss_e}. The
transition between the shocked and unshocked regions, is obtained
by Rankine--Hugoniot jump conditions. In this section we will derive
expressions for these jump conditions in terms of the values of the
similarity profiles at the shock surface. 

First, we note that the general form of the Rankine--Hugoniot jump
conditions for a set of conserved quantities $\boldsymbol{u}\left(y,t\right)$
which obey a set of (nonlinear) hyperbolic conservation laws in the
form $\frac{\partial\boldsymbol{u}}{\partial t}+\frac{\partial}{\partial y}\boldsymbol{F}\left(\boldsymbol{u}\right)=0$,
are written as $\dot{y}\Delta\boldsymbol{u}=\Delta\boldsymbol{F}$,
where $\dot{y}$ is the local discontinuity velocity and $\Delta\boldsymbol{u},\Delta\boldsymbol{F}$
are, respectively, the differences in the conserved quantities $\boldsymbol{u}$
and fluxes $\boldsymbol{F}\left(\boldsymbol{u}\right)$ across the
discontinuity \cite{leveque2002finite,toro2013riemann}. We want to
derive the jump conditions for the hydrodynamic equations \ref{eq:_mass_hydro}-\ref{eq:energy_hydro},
which are written in Lagrangian form. We note that the internal energy
equation \ref{eq:energy_hydro}, is not written in conservative form.
Instead, we use the equation for the conservation of total energy:
\begin{equation}
\frac{\partial}{\partial t}\left(\frac{pv}{r}+\frac{1}{2}u^{2}\right)+\frac{\partial\left(pu\right)}{\partial m}=0,\label{eq:etot}
\end{equation}
which is in conservative form. The Rankine--Hugoniot jump conditions
for the conservation of mass (equation \ref{eq:_mass_hydro}), momentum
(equation \ref{eq:p_hydro}) and total energy (equation \ref{eq:etot})
are given by:

\begin{equation}
\dot{m}\left(v_{s}-v_{\text{ref}}\right)=u_{\text{ref}}-u_{s},
\end{equation}
\begin{align}
\dot{m}\left(u_{s}-u_{\text{ref}}\right) & =p_{s}-p_{\text{ref}},
\end{align}

\begin{align}
 & \dot{m}\left[\left(\frac{p_{s}v_{s}}{r}+\frac{1}{2}u_{s}^{2}\right)-\left(\frac{p_{\text{ref}}v_{\text{ref}}}{r}+\frac{1}{2}u_{\text{ref}}^{2}\right)\right]\nonumber \\
 & =u_{s}p_{s}-u_{\text{ref}}p_{\text{ref}},
\end{align}
where 'ref' represents quantities in the (reference) unshocked region
and 's' represents quantities in the shocked region. $\dot{m}$ is
the shock mass flux (the shock velocity in mass space). Since in the
unshocked region $u_{\text{ref}}=p_{\text{ref}}=0$, the jump conditions
describe a strong shock \cite{zeldovich1967physics,mihalas2013foundations}:
\begin{equation}
\dot{m}\left(v_{s}-v_{\text{ref}}\right)=-u_{s},\label{eq:HUG_MASS}
\end{equation}
\begin{align}
\dot{m}u_{s} & =p_{s},\label{eq:HUG_MOM}
\end{align}

\begin{equation}
\dot{m}\left(\frac{p_{s}v_{s}}{r}+\frac{1}{2}u_{s}^{2}\right)=u_{s}p_{s}.\label{eq:HUG_E}
\end{equation}
In order to write these equations in terms of the similarity profiles,
we note that the shocked mass and mass flux are given by:
\begin{equation}
m_{s}\left(t\right)=v_{0}^{-\frac{1}{2-\omega}}p_{0}^{\frac{1-\omega}{2-\omega}}t^{\frac{\left(2+\tau\right)\left(1-\omega\right)}{2-\omega}}\xi_{s},\label{eq:shocked_mass}
\end{equation}

\begin{equation}
\dot{m}=v_{0}^{-\frac{1}{2-\omega}}p_{0}^{\frac{1-\omega}{2-\omega}}t^{\frac{\left(2+\tau\right)\left(1-\omega\right)}{2-\omega}-1}\left(2+\tau\right)\left(\frac{1-\omega}{2-\omega}\right)\xi_{s},\label{eq:mdot}
\end{equation}
where $\xi_{s}$ is the similarity variable at the shock front, which
is not known at this point. The temporal similarity exponent for the
mass, $\frac{\left(2+\tau\right)\left(1-\omega\right)}{2-\omega}$,
is plotted in Figure \ref{fig:mass_pow} as a function of $\tau,\omega$.
We see that the requirement for the shock to propagate, gives the
following constraint on the piston pressure temporal exponent:
\begin{equation}
\tau>-2.
\end{equation}
The specific volume in the unshocked side is given by (see equation
\ref{eq:init_conds_v_lag}):
\begin{align}
v_{\text{ref}} & =\left(1-\omega\right)^{\frac{\omega}{1-\omega}}v_{0}^{\frac{2}{2-\omega}}p_{0}^{\frac{\omega}{2-\omega}}t^{\frac{\left(2+\tau\right)\omega}{2-\omega}}\xi_{s}^{\frac{\omega}{1-\omega}},\label{eq:vref}
\end{align}
while in the shocked side:
\begin{equation}
v_{s}=V_{s}v_{0}^{\frac{2}{2-\omega}}p_{0}^{\frac{\omega}{2-\omega}}t^{\frac{\left(2+\tau\right)\omega}{2-\omega}},\label{eq:vs}
\end{equation}

\begin{equation}
u_{s}=U_{s}v_{0}^{\frac{1}{2-\omega}}p_{0}^{\frac{1}{2-\omega}}t^{\frac{\omega+\tau}{2-\omega}},\label{eq:us}
\end{equation}
\begin{equation}
p_{s}=P_{s}p_{0}t^{\tau},\label{eq:ps}
\end{equation}
where we have defined $V_{s}=V\left(\xi_{s}\right),U_{s}=U\left(\xi_{s}\right),P_{s}=P\left(\xi_{s}\right)$.
By substituting equations \ref{eq:mdot}-\ref{eq:ps} in the jump
conditions \ref{eq:HUG_MASS}-\ref{eq:HUG_E}, all dimensional terms
factor out, and the following dimensionless jump conditions are obtained:
\begin{equation}
V_{s}=\left(\left(1-\omega\right)\xi_{s}\right)^{\frac{\omega}{1-\omega}}-\frac{U_{s}}{\left(2+\tau\right)\left(\frac{1-\omega}{2-\omega}\right)\xi_{s}},
\end{equation}

\begin{equation}
P_{s}=\left(2+\tau\right)\left(\frac{1-\omega}{2-\omega}\right)\xi_{s}U_{s},
\end{equation}

\begin{equation}
\left(2+\tau\right)\left(\frac{1-\omega}{2-\omega}\right)\xi_{s}\left(\frac{P_{s}V_{s}}{r}+\frac{1}{2}U_{s}^{2}\right)=U_{s}P_{s}.\label{eq:hug_etot_ss}
\end{equation}
Solving these equations for $V_{s},U_{s},P_{s}$ gives the following
expressions for the similarity profiles at the shock in terms of $\xi_{s}$:
\begin{equation}
V_{s}=\frac{r}{r+2}\left(\left(1-\omega\right)\xi_{s}\right)^{\frac{\omega}{1-\omega}},\label{eq:vs_hug}
\end{equation}
\begin{equation}
U_{s}=\frac{2}{r}\left(2+\tau\right)\left(\frac{1-\omega}{2-\omega}\right)\xi_{s}V_{s},\label{eq:us_hug}
\end{equation}

\begin{equation}
P_{s}=\frac{U_{s}^{2}}{2}\frac{r}{V_{s}}.\label{eq:ps_hug}
\end{equation}
As before, we see that for $\omega=0$, these equations are reduced
to equations 46a-46d in Ref. \cite{shussman2015full}.

\section{Solution of the similarity equations}

\begin{figure}[t]
\begin{centering}
\includegraphics[scale=0.55]{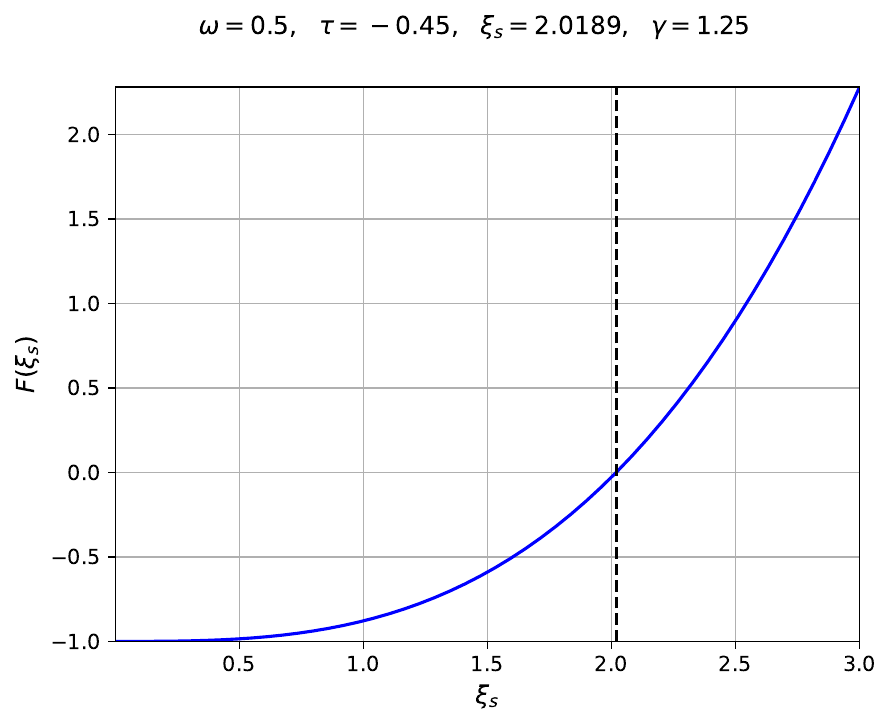}
\par\end{centering}
\caption{The function $F\left(\xi_{s}\right)=P_{\xi_{s}}\left(0\right)-1$
which has a root at the correct value of the similarity variable at
the shock, $\xi_{s}$. The calculation was performed for $\omega=0.5$,
$\tau=-0.45$ and $\gamma=1.25$, resulting in $\xi_{s}=2.0189$ (as
shown by the black dashed vertical line). \label{fig:fxsi_s}}
\end{figure}

\begin{figure}[t]
\begin{centering}
\includegraphics[scale=0.6]{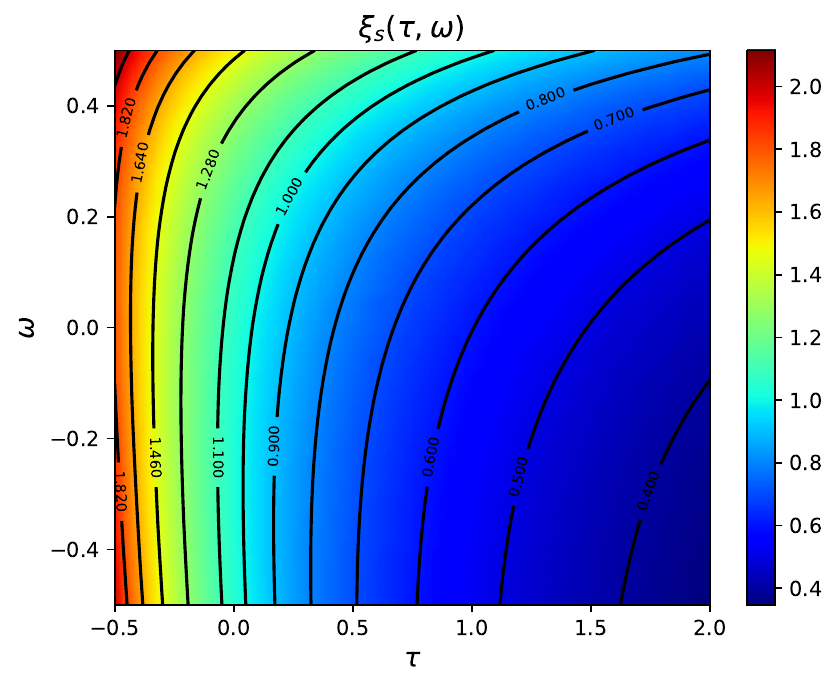}
\par\end{centering}
\caption{A color plot for the shock front similarity variable $\xi_{s}$ as
a function of the piston pressure temporal power $\tau$ and the initial
density spatial power $\omega$ and an adiabatic index $\gamma=1.25$.\label{fig:xsi_s_tau}}
\end{figure}

\begin{figure}[t]
\begin{centering}
\includegraphics[scale=0.55]{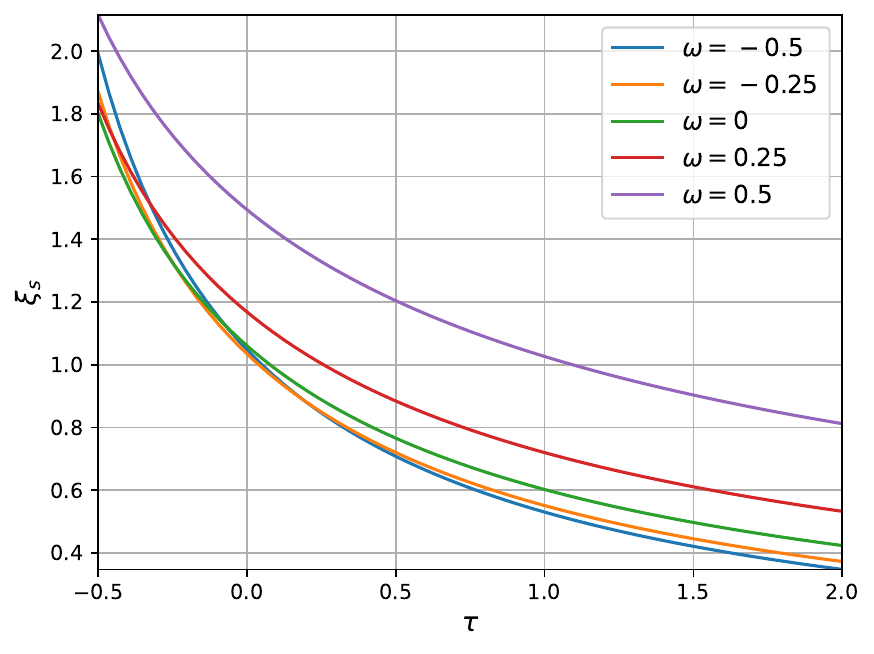}
\par\end{centering}
\caption{Shock front similarity variable $\xi_{s}$ as a function of $\tau$
for selected values of $\omega$ and as a function of $\tau,\omega$
and $\gamma=1.25$. \label{fig:xsi_s_tau-1}}
\end{figure}

\begin{figure*}[t]
\begin{centering}
\includegraphics[scale=0.38]{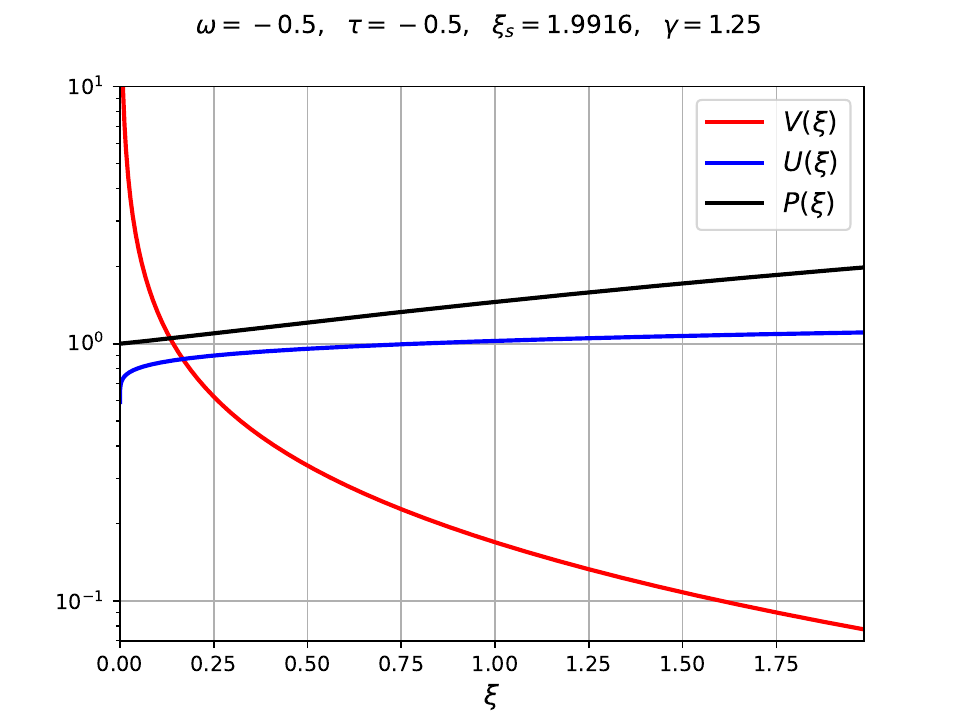}\includegraphics[scale=0.38]{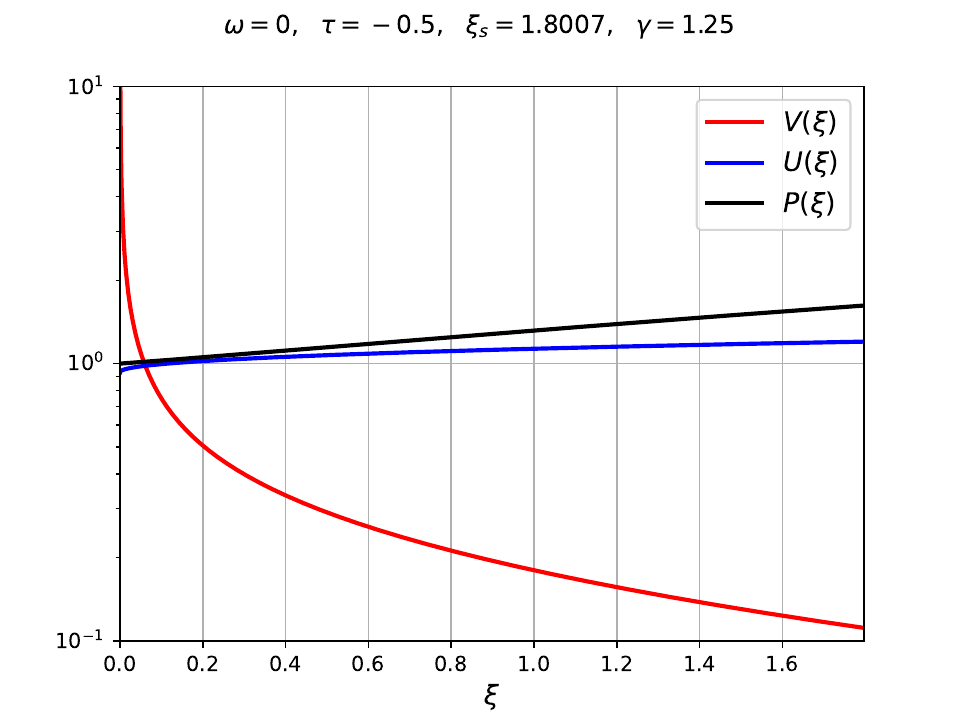}\includegraphics[scale=0.38]{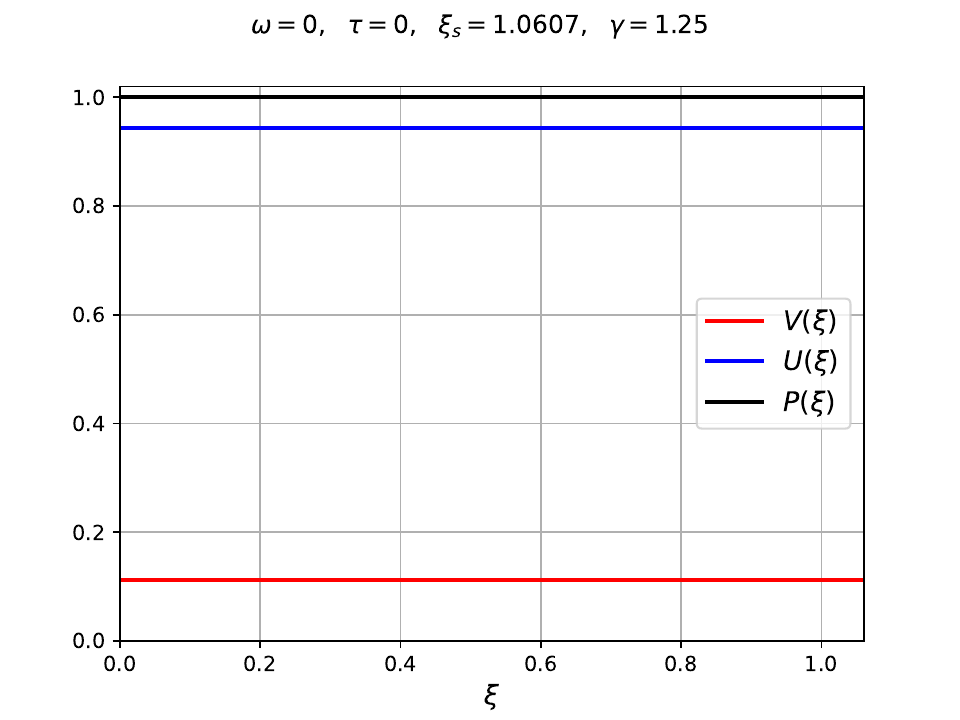}
\par\end{centering}
\begin{centering}
\includegraphics[scale=0.38]{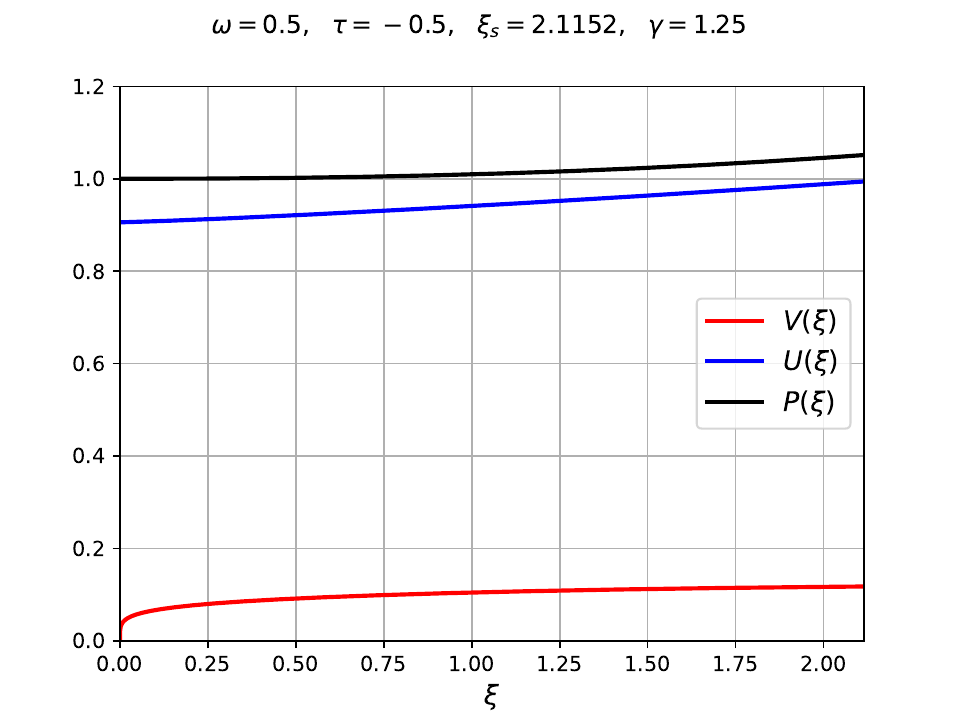}\includegraphics[scale=0.38]{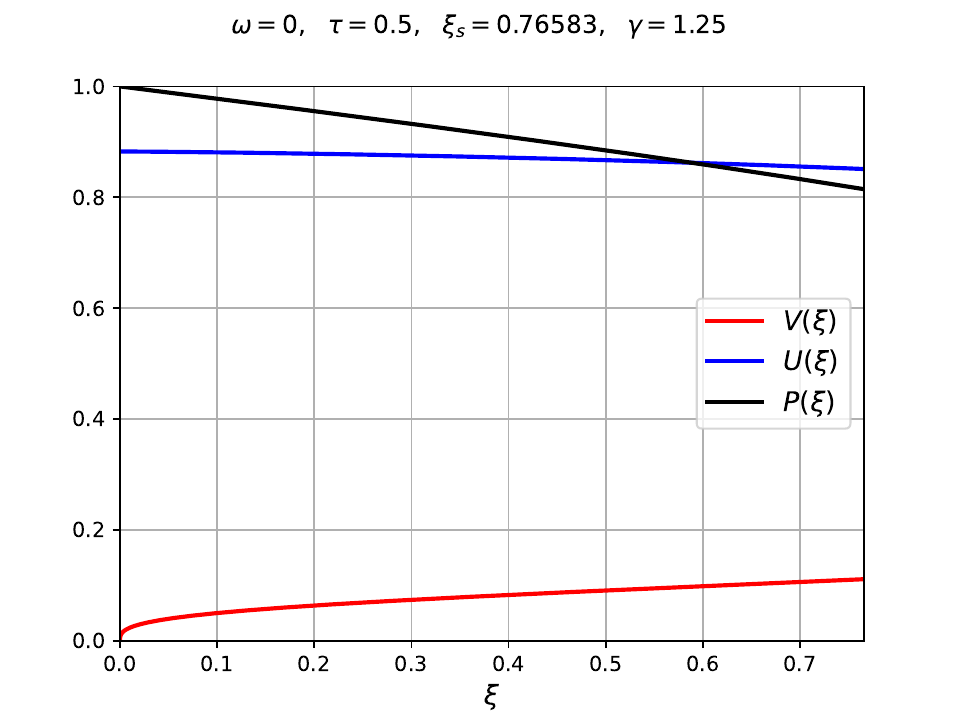}\includegraphics[scale=0.38]{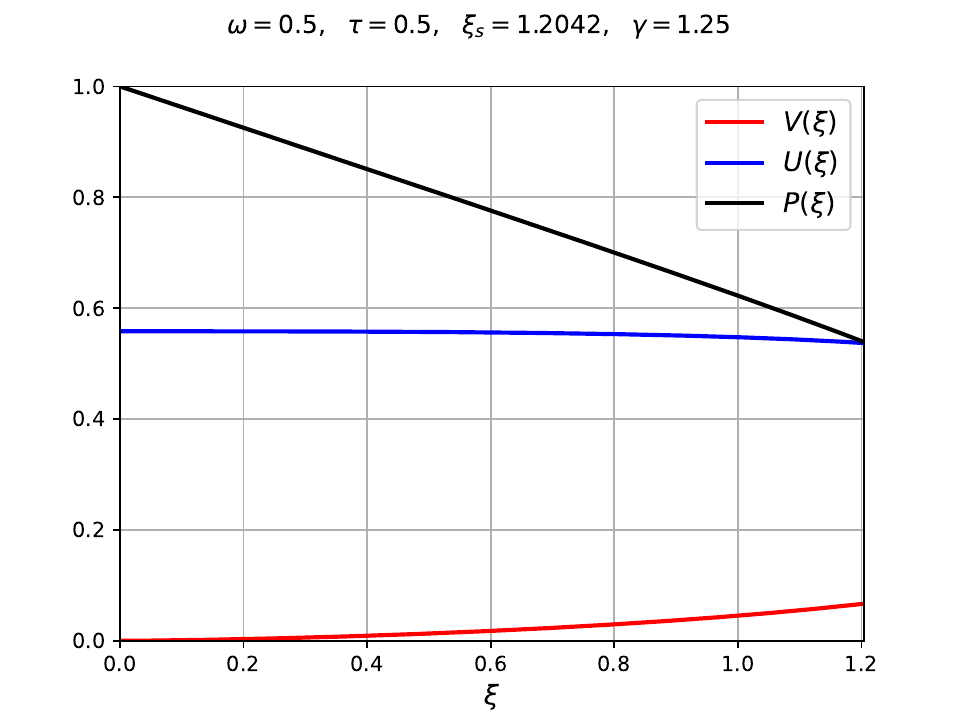}
\par\end{centering}
\caption{Similarity profiles of the specific volume $V\left(\xi\right)$ (red),
velocity $U\left(\xi\right)$ (blue) and pressure $P\left(\xi\right)$
(black), for various values of $\tau,\omega$ and $\gamma=1.25$,
as listed in the title of each sub-plot. \label{fig:ss_prof}}
\end{figure*}

\begin{figure}[t]
\begin{centering}
\includegraphics[scale=0.6]{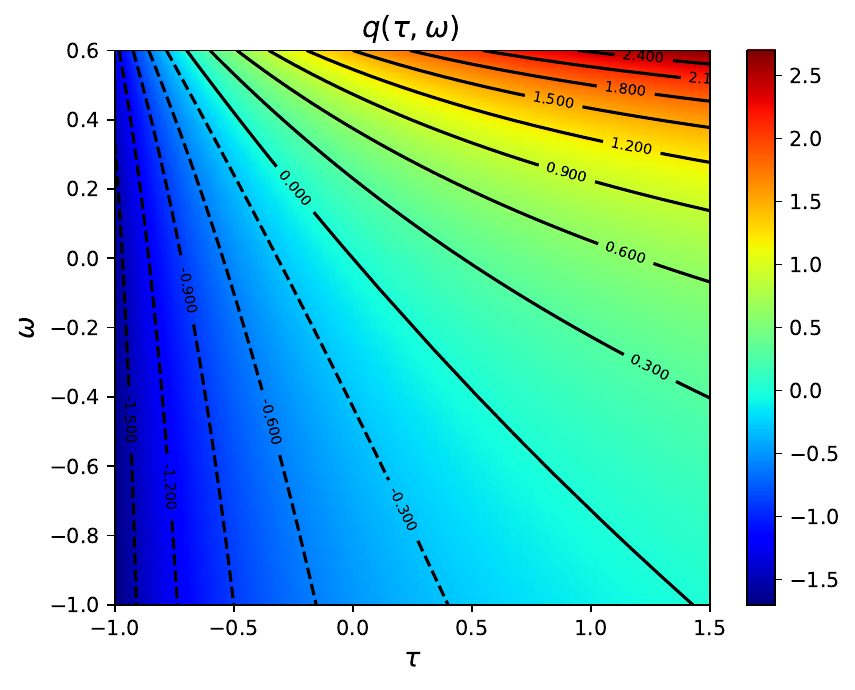}
\par\end{centering}
\caption{A color plot for the asymptotic power $q$ (see equation \ref{eq:qdef})
of the specific volume which behaves as $V\left(\xi\right)\approx A\xi^{q}$
near the piston, as a function of $\tau,\omega$ and $\gamma=1.25$.
The line $q\left(\tau,\omega\right)=0$ (given in equation \ref{eq:omega_tau_q0})
represents a family of solutions with a finite density near the piston,
and divides the $\tau-\omega$ plane into regions for which the density
near the piston diverges, ($q>0)$ or vanishes ($q<0$).\label{fig:qpow}}
\end{figure}

In order to solve the ordinary differential equations (ODE) \ref{eq:eq_ss_mass}-\ref{eq:eq_ss_e}
for the similarity profiles, we write them in the canonical form $\boldsymbol{y}'=\boldsymbol{f}(\boldsymbol{y})$
by employing Kramers' law:

\begin{equation}
V'=\frac{\Delta_{1}}{\Delta},\ U'=\frac{\Delta_{2}}{\Delta},\ P'=\frac{\Delta_{3}}{\Delta},\label{eq:canonical_ode_form}
\end{equation}
where:
\begin{equation}
\Delta\left(V,P\right)=a^{3}\xi^{2}V-bP,\label{eq:delta}
\end{equation}
\begin{equation}
\Delta_{1}\left(V,U,P\right)=-V\left(c_{1}a^{2}\xi V+c_{2}aU+\frac{c_{3}}{\xi}P\right),\label{eq:D1}
\end{equation}

\begin{equation}
\Delta_{2}\left(V,U,P\right)=V\left(c_{1}bP+c_{2}a^{2}\xi U+c_{3}aP\right),\label{eq:D2}
\end{equation}
\begin{equation}
\Delta_{3}\left(V,U,P\right)=P\left(c_{1}ab\xi V+c_{2}bU+c_{3}a^{2}\xi V\right),\label{eq:D3}
\end{equation}
with the following coefficients:
\begin{equation}
a=\frac{\left(2+\tau\right)\left(1-\omega\right)}{2-\omega},\ b=\left(1+r\right)a,
\end{equation}
\begin{equation}
c_{1}=-\frac{\omega\left(2+\tau\right)}{2-\omega},\ c_{2}=\frac{\omega+\tau}{2-\omega},\label{eq:C12}
\end{equation}
\begin{equation}
c_{3}=\tau+\left(1+r\right)\frac{\left(2+\tau\right)\omega}{2-\omega}.\label{eq:C3}
\end{equation}
The initial conditions for equation \ref{eq:canonical_ode_form} are
the jump conditions \ref{eq:vs_hug}-\ref{eq:ps_hug} at $\xi=\xi_{s}$.
The piston boundary condition \ref{eq:p_bc_ss} gives a constraint
from which the value of $\xi_{s}$ is determined. This can be performed
numerically by finding the root of the function $F\left(\xi_{s}\right)=P_{\xi_{s}}\left(0\right)-1$
where $P_{\xi_{s}}\left(0\right)$ is the similarity pressure profile
at $\xi=0$ that was obtained by a numerical integration of the ODE
\ref{eq:canonical_ode_form} from a trial value $\xi=\xi_{s}$ to
$\xi=0$. An example of the root finding procedure for the calculation
of $\xi_{s}$ is presented in Fig. \ref{fig:fxsi_s}.

In Figures \ref{fig:xsi_s_tau}-\ref{fig:xsi_s_tau-1} we present
calculations of $\xi_{s}$ as a function of $\tau,\omega$ in a wide
range. In Figure \ref{fig:ss_prof} the similarity profiles $V\left(\xi\right)$,
$U\left(\xi\right)$, $P\left(\xi\right)$ are plotted for various
values of $\tau,\omega$. It is seen that for all cases shown, except
the special case $\tau=\omega=0$ (a constant piston pressure and
a uniform initial density), the specific volume either diverges or
vanishes near $\xi=0$ (the location of the piston). This behavior
can be characterized by performing an asymptotic analysis near $\xi\rightarrow0$,
of equation \ref{eq:canonical_ode_form} for the specific volume.
Using the fact that $U\left(\xi\right)$, $P\left(\xi\right)$ are
finite near $\xi\rightarrow0$ and, as will be shown subsequently
in section \ref{sec:x-t}, we also have$\lim_{\xi\rightarrow0}\xi V\left(\xi\right)=0$,
the equation for the specific volume near $\xi\rightarrow0$ can be
written as:
\begin{equation}
V'=q\frac{V}{\xi},
\end{equation}
where:
\begin{equation}
q\left(\omega,\tau\right)=\frac{c_{3}}{b}=\frac{\tau\left(2-\omega\right)}{\left(r+1\right)\left(2+\tau\right)\left(1-\omega\right)}+\frac{\omega}{1-\omega}.\label{eq:qdef}
\end{equation}
This equation has a power law solution of the form $V\left(\xi\right)=A\xi^{q}$.
This asymptotic power is plotted in Figure \ref{fig:qpow} for a wide
range of $\tau,\omega$. The line $q\left(\tau,\omega\right)=0$,
which is given by the relation:
\begin{equation}
\omega\left(\tau\right)=\frac{2\tau}{\tau-\left(r+1\right)\left(2+\tau\right)},\label{eq:omega_tau_q0}
\end{equation}
 represents a family of solutions which have a finite density near
the piston, and divides the $\tau-\omega$ plane into regions for
which the density near the piston diverges ($q>0)$, or vanishes ($q<0$). 

We note that for the trivial case $\tau=\omega=0$, we have $q=0$,
and the density is finite near the piston. In this case, the coefficients
in equations \ref{eq:C12}-\ref{eq:C3} are zero, which results in
the vanishing of the determinants \ref{eq:D1}-\ref{eq:D3}, so that
the ODE \ref{eq:canonical_ode_form} has a trivial analytical solution
for which the similarity profiles are constant (as also seen in Figure
\ref{fig:ss_prof}). Since the pressure at the piston ($\xi=0$) is
known from the boundary condition \ref{eq:p_bc_ss}, the value of
$V,U$ and $\xi_{s}$ can be obtained analytically by inverting the
strong shock jump conditions \ref{eq:vs_hug}-\ref{eq:ps_hug}, resulting
in the trivial solution:
\begin{equation}
\xi_{s}=\left(\frac{r+2}{2}\right)^{\frac{1}{2}},
\end{equation}
\begin{equation}
V\left(\xi\right)=\frac{r}{r+2},
\end{equation}
\begin{equation}
U\left(\xi\right)=\left(\frac{2}{r+2}\right)^{\frac{1}{2}},
\end{equation}
\begin{equation}
P\left(\xi\right)=1.
\end{equation}
For other values of $\tau,\omega$ for which $q=0$ (according to
equation \ref{eq:omega_tau_q0}), even though the density is finite
near the origin, the determinants \ref{eq:D1}-\ref{eq:D3} are nonzero,
and the solution is not constant, so a numerical procedure for the
determination of $\xi_{s}$ , as described above is necessary. In
summary, other than the trivial solution for $\tau=\omega=0$, we
are not aware of any closed form analytical solutions of equation
\ref{eq:canonical_ode_form}, which therefore must be solved numerically,
yielding a semi-analytical self similar solution.

Finally, we note that the ODE in the form \ref{eq:canonical_ode_form}
has a potential singularity if $\Delta\left(V,P\right)=0$. The emergence
of singularities is a typical behavior which appears in the construction
of similarity solutions of the second kind \cite{barenblatt1996scaling,zeldovich1967physics},
and specifically in the work on converging shock waves \cite{lazarus1981self,chisnell1998analytic,ramsey2012guderley,modelevsky2021revisiting,giron2021solutions}.
Regarding the planar piston driven shock wave solution, if the determinant
in equation \ref{eq:delta} is zero, we must have the relation $P\left(\xi\right)=a^{2}\xi^{2}V\left(\xi\right)/\left(1+r\right)$.
This relation results in a vanishing pressure near the piston ($\xi=0$),
and therefore cannot hold there, since the pressure is finite according
to the piston boundary condition \ref{eq:p_bc_ss}. Moreover, if a
singular behavior existed in some finite range $0<\xi_{1}<\xi<\text{\ensuremath{\xi_{2}}}$,
this relation can be inserted to the energy ODE \ref{eq:eq_ss_e},
resulting in a non-singular power law solution of the form $V\left(\xi\right)=B\xi^{l}$,
where $B$ is a constant and $l=\frac{1}{r}\left(2+\frac{\tau}{a}\right)+\left(1+\frac{1}{r}\right)\frac{\omega}{1-\omega}$.
In practice, we did not encounter any singular behavior in the numerical
integration of eq. \ref{eq:canonical_ode_form}, for a wide range
of $\tau,\omega$ (see results in figures \ref{fig:xsi_s_tau},\ref{fig:umesh},\ref{fig:erat}),
which shows that the solutions did not pass through any singular points.
We conclude that the piston driven shock wave equations \ref{eq:canonical_ode_form}
do not seem to posses any singularities, which is a common property
of self-similar solutions of the first kind, such as the Noh stagnation
shock or Sedov-Taylor-von Neumann blast wave problems.

\section{The solution in real space\label{sec:x-t}}

In order to write the solution in real space (that is, in Eulerian
coordinates), we will now derive exact an expression for the temporal
evolution of the positions of Lagrangian mass elements. For an element
$m>m_{s}\left(t\right)$ which the shock did not reach at time $t$,
the position is given by the initial position (via equation \ref{eq:position_mnass_initial}).
For a shocked mass element $m<m_{s}\left(t\right)$, we have:
\begin{equation}
x\left(m,t\right)-x_{p}\left(t\right)=\int_{0}^{m}v\left(m',t\right)dm',
\end{equation}
where $x_{p}\left(t\right)$ is the piston position. By substituting
the similarity volume profile (equation \ref{eq:v_ss}), one obtains
the relation:
\begin{align}
x\left(m,t\right)-x_{p}\left(t\right) & =\left(v_{0}p_{0}t^{2+\tau}\right)^{\frac{1}{2-\omega}}\int_{0}^{\xi}V\left(\xi'\right)d\xi',\label{eq:xm1}
\end{align}
where the upper integration limit $\xi=\xi\left(m\right)$ is given
from equation \ref{eq:xsi_def}. We note that the integral in equation
\ref{eq:xm1} can be calculated analytically, by integrating equation
\ref{eq:eq_ss_mass} one finds:
\begin{equation}
\int_{0}^{\xi}V\left(\xi'\right)d\xi'=\left(1-\omega\right)\xi V\left(\xi\right)+\left(\frac{2-\omega}{2+\tau}\right)\left(U\left(\xi\right)-U\left(0\right)\right),\label{eq:vint}
\end{equation}
where we have used the fact that in order for the integral $\int_{0}^{\xi}V\left(\xi'\right)d\xi'$,
which represents position, to be finite, we must have $\lim_{\xi\rightarrow0}\left[V\left(\xi\right)\xi\right]=0$. 

\begin{figure}[t]
\begin{centering}
\includegraphics[scale=0.6]{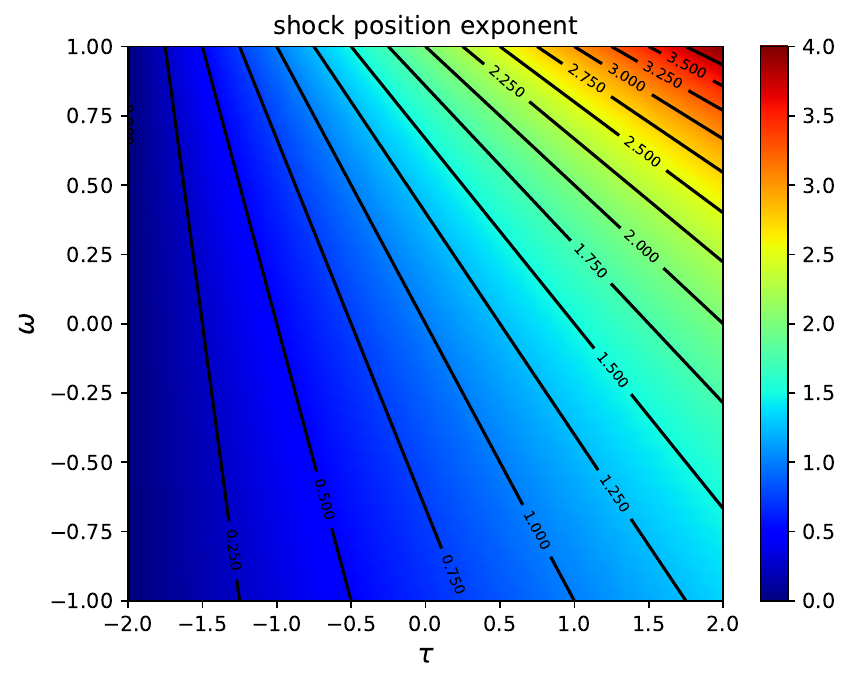}
\par\end{centering}
\caption{A color plot for the position similarity exponent $\frac{2+\tau}{2-\omega}$
(which is the temporal power of the position of any mass element,
including the shock and piston positions, see text), as a function
of $\tau,\omega$. The line $\frac{2+\tau}{2-\omega}=1$ represents
solutions with a constant shock speed, and divides the $\tau-\omega$
plane into regions of solutions with accelerating (above the line)
or decelerating (below the line) shocks.\label{fig:xpow}}
\end{figure}

The piston position is obtained by a temporal integration of the piston
velocity $u_{p}\left(t\right)\equiv u\left(m=0,t\right)$, and using
equation \ref{eq:u_ss}:
\begin{align}
x_{p}\left(t\right) & =v_{0}^{\frac{1}{2-\omega}}p_{0}^{\frac{1}{2-\omega}}t^{\frac{2+\tau}{2-\omega}}\left(\frac{2-\omega}{2+\tau}\right)U\left(0\right).\label{eq:xpiston}
\end{align}
In summary, the position of a Lagrangian element $m$ as a function
of time is given by the following formula:
\begin{align}
 & x\left(m,t\right)=\label{eq:xmt}\\
 & \begin{cases}
\left[v_{0}\left(1-\omega\right)m\right]^{\frac{1}{1-\omega}},\ \ \ \ \ \ \ \ \ \ \ \ \ \ \ \ \ \ \ \ \ \ \ \ \ \ \ \ \ \ \ m>m_{s}\left(t\right)\\
\left(v_{0}p_{0}t^{2+\tau}\right)^{\frac{1}{2-\omega}}\left[\left(1-\omega\right)\xi V\left(\xi\right)+\left(\frac{2-\omega}{2+\tau}\right)U\left(\xi\right)\right],\ \text{else}
\end{cases}\nonumber 
\end{align}
We note that for the shock position $x_{s}\left(t\right)\equiv x\left(m_{s}\left(t\right),t\right)$,
both expressions in equation \ref{eq:xmt} coincide, since $\xi=\xi_{s}$
and due to the jump conditions \ref{eq:vs_hug}-\ref{eq:us_hug}.
This gives a simple expression for the shock position in term of $\xi_{s}$:
\begin{align}
x_{s}\left(t\right) & =\left(v_{0}p_{0}t^{2+\tau}\right)^{\frac{1}{2-\omega}}\left(\left(1-\omega\right)\xi_{s}\right)^{\frac{1}{1-\omega}},\label{eq:xshock}
\end{align}
 The position temporal exponent $\frac{2+\tau}{2-\omega}$, which
governs the temporal behavior of any mass element (and specifically
the shock and piston position), is plotted in Figure \ref{fig:xpow}
as a function of $\tau,\omega$. The line $\frac{2+\tau}{2-\omega}=0$
represents a family of solutions with a constant shock velocity, and
divides the $\tau-\omega$ plane into regions for which the shock
accelerates or slows down over time.

\begin{figure}[t]
\begin{centering}
\includegraphics[scale=0.6]{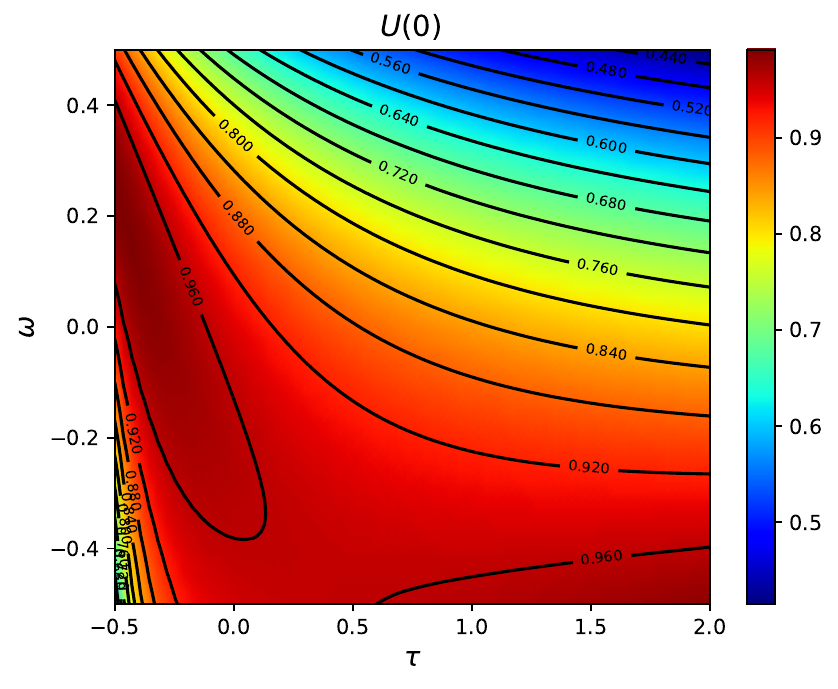}
\par\end{centering}
\caption{A color plot for the the similarity piston velocity $U\left(0\right)$,
as a function of $\tau,\omega$ for $\gamma=1.25$.\label{fig:umesh}}
\end{figure}

\begin{figure}[t]
\begin{centering}
\includegraphics[scale=0.55]{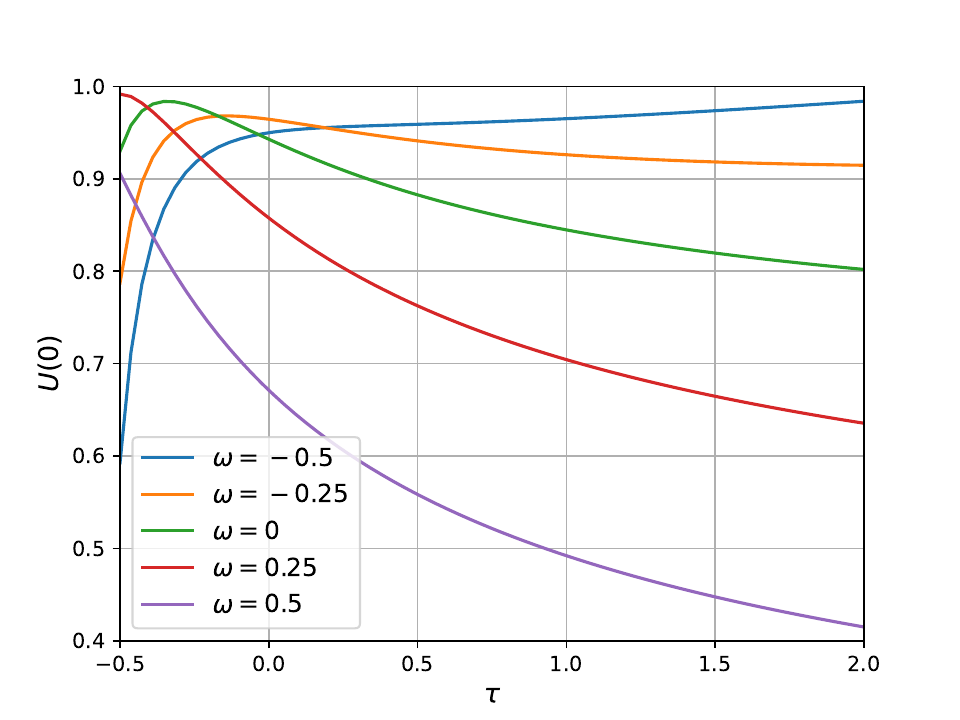}
\par\end{centering}
\caption{The similarity piston velocity, $U\left(0\right)$, as a function
of $\tau$ for selected values of $\omega$ and $\gamma=1.25$. \label{fig:utau}}
\end{figure}

The quantity $U\left(0\right)$, which sets the piston velocity, is
plotted as a function of $\tau,\omega$ in Figures \ref{fig:umesh}-\ref{fig:utau}.
It was obtained from a numerical integration of the similarity ODE,
as described in the previous section.

\begin{figure*}[t]
\begin{centering}
\includegraphics[scale=0.38]{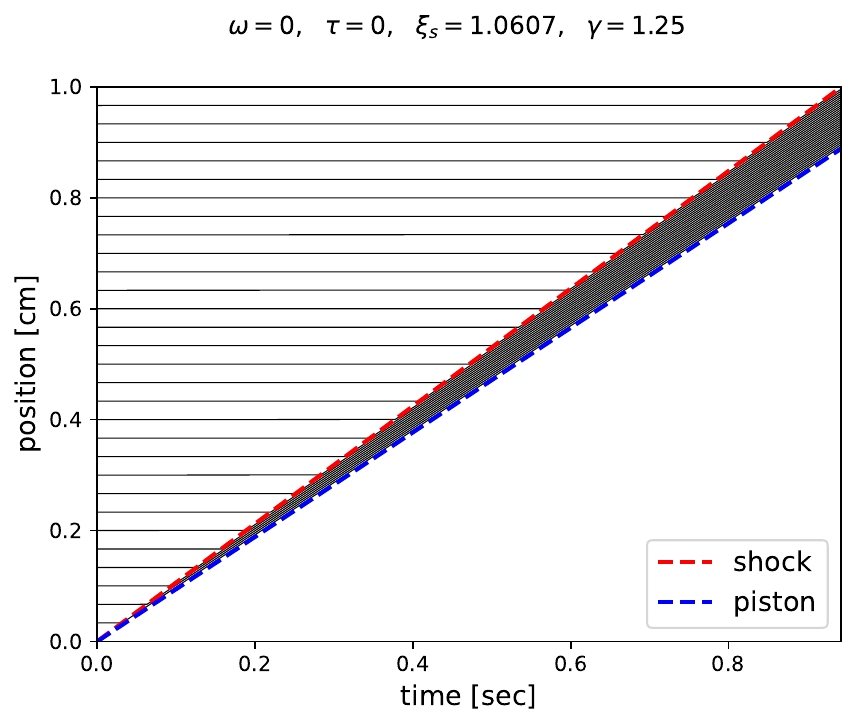}\includegraphics[scale=0.38]{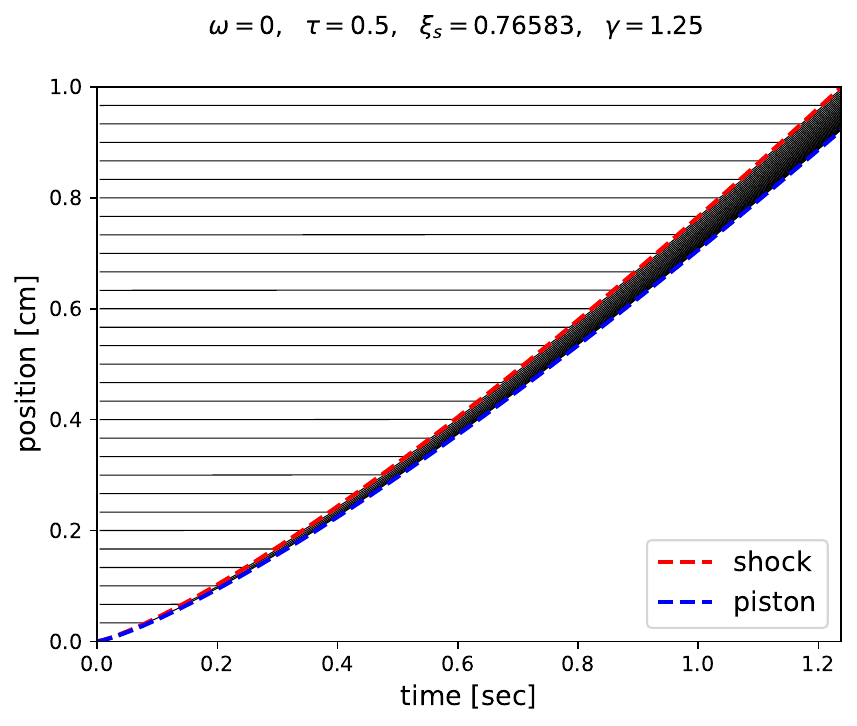}\includegraphics[scale=0.38]{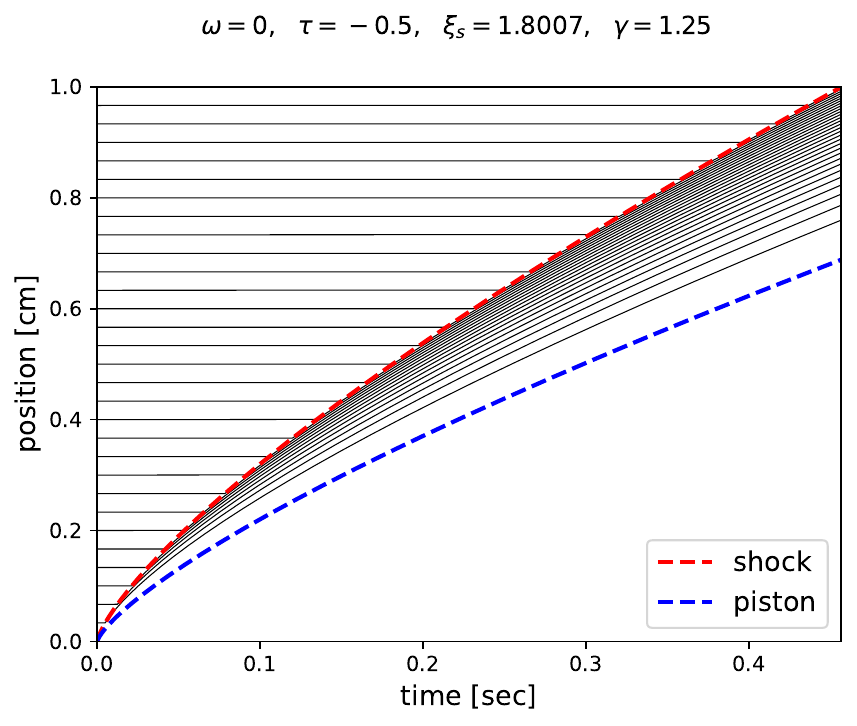}
\par\end{centering}
\begin{centering}
\includegraphics[scale=0.38]{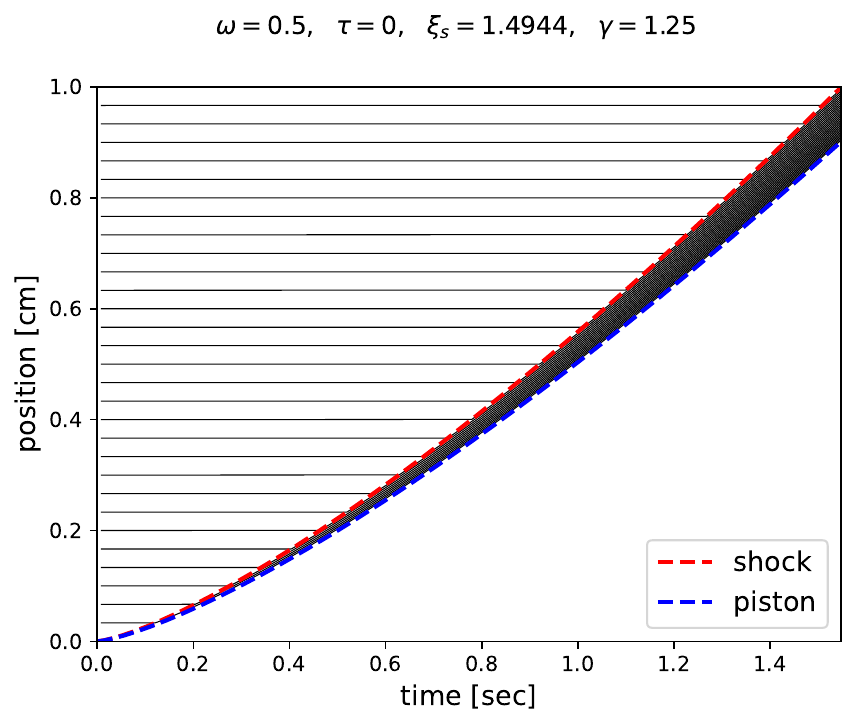}\includegraphics[scale=0.38]{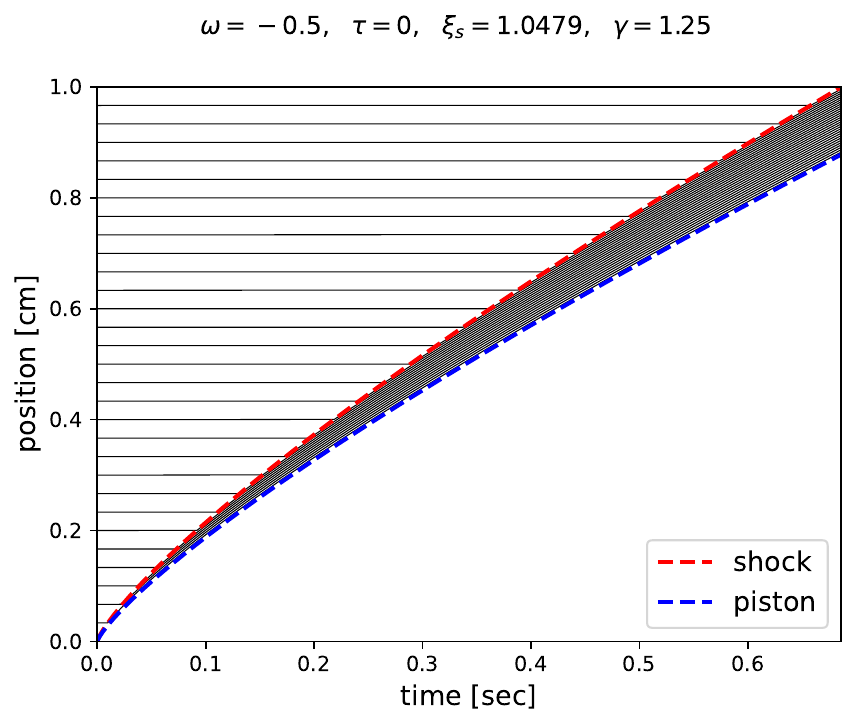}\includegraphics[scale=0.38]{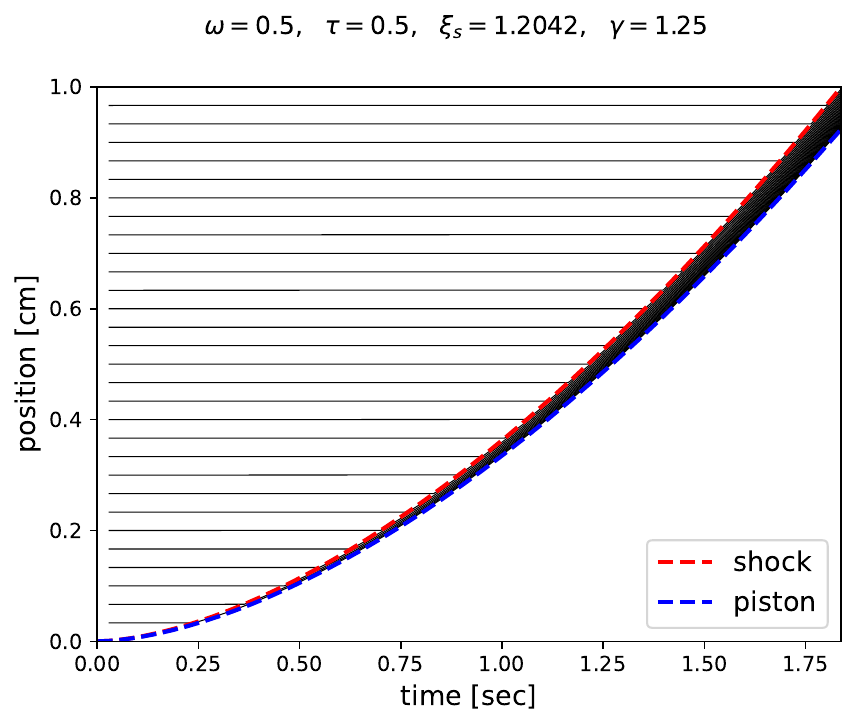}
\par\end{centering}
\begin{centering}
\includegraphics[scale=0.38]{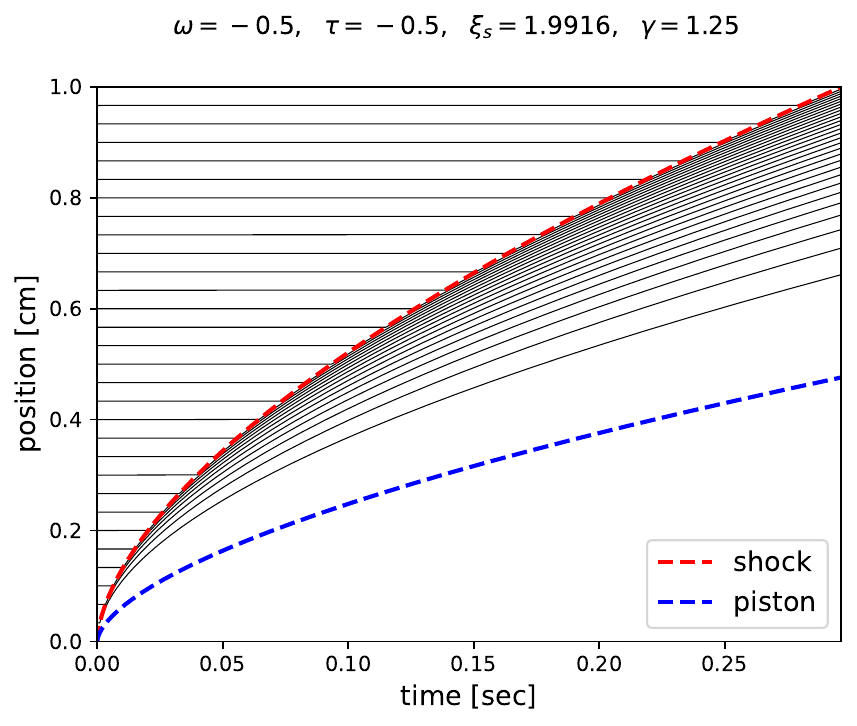}\includegraphics[scale=0.38]{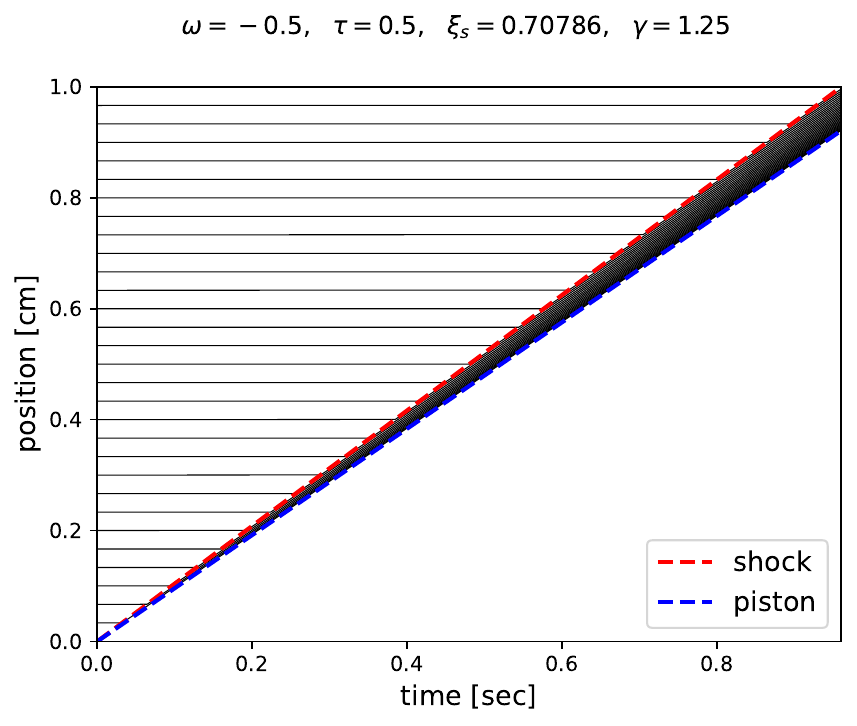}\includegraphics[scale=0.38]{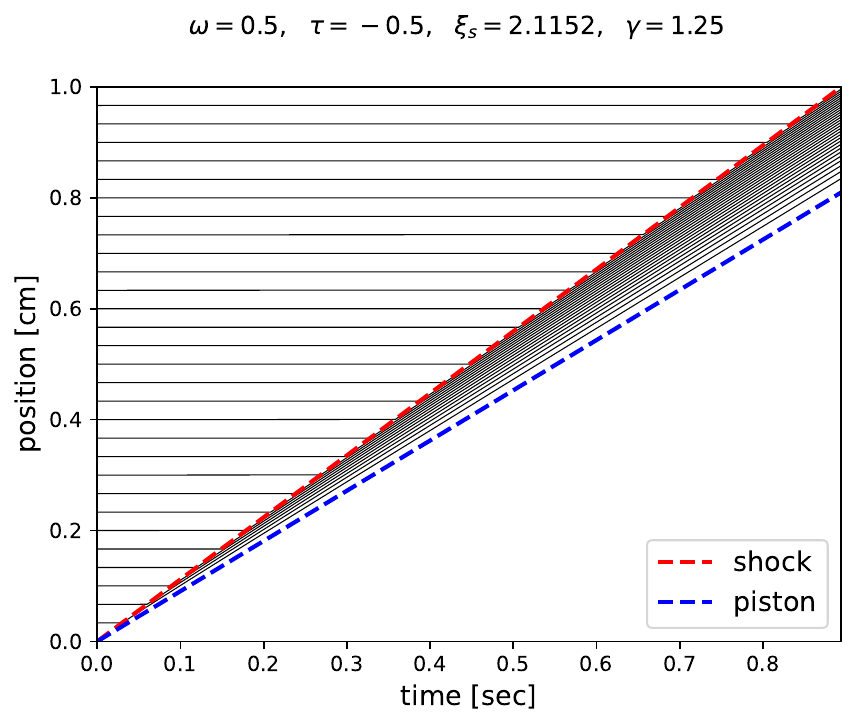}
\par\end{centering}
\begin{centering}
\includegraphics[scale=0.38]{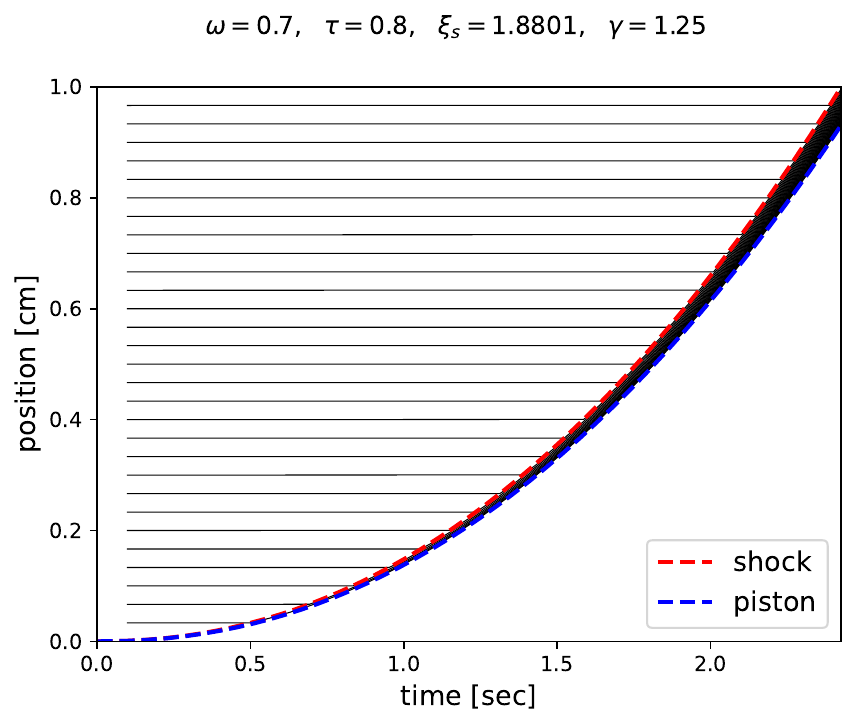}\includegraphics[scale=0.38]{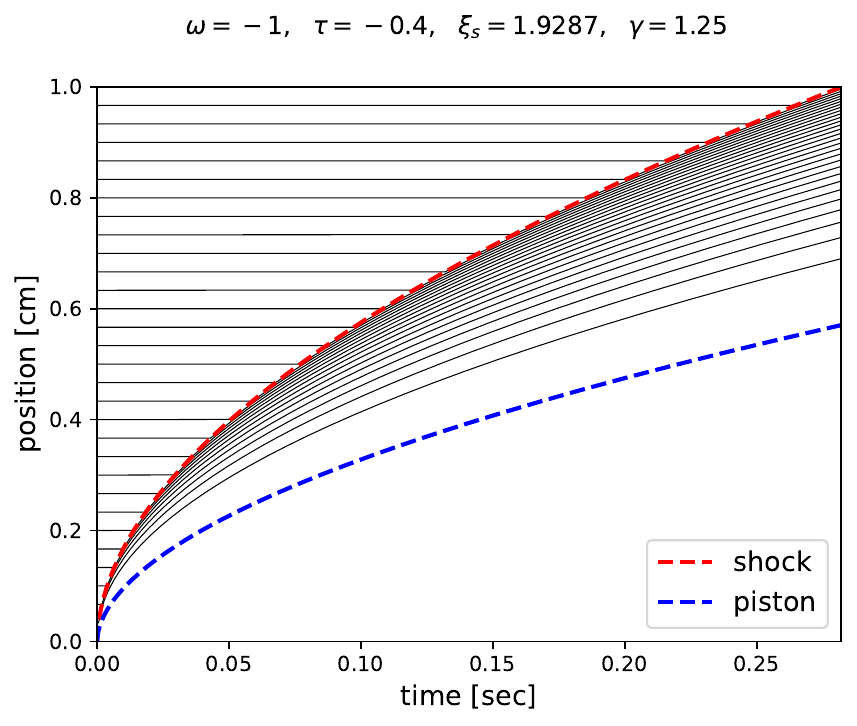}\includegraphics[scale=0.38]{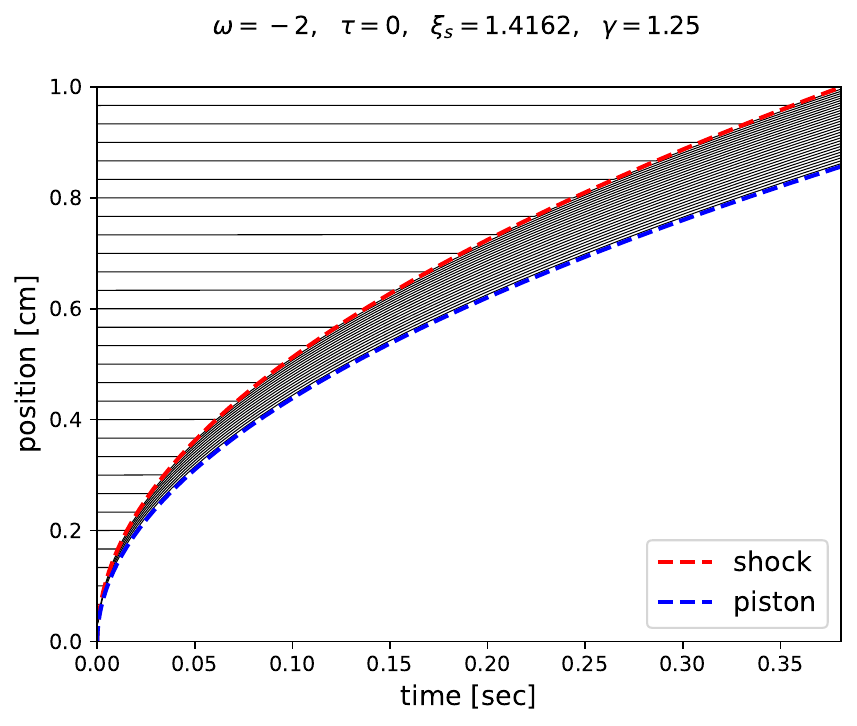}
\par\end{centering}
\caption{$x-t$ diagrams for $\gamma=1.25$, $\rho_{0}=1\text{g/cm}$, $p_{0}=1\text{dyn/cm}^{2}/\text{sec}^{\tau}$
and various values of $\tau,\omega$, as listed in the title of each
subplot. The locations of the piston (blue) and the shock (red) are
also presented. The diagrams describe the exact locations of Lagrangian
mass elements (via equation \ref{eq:xmt}), through which the shock
is propagating. It is evident that different values of $\tau,\omega$
result in constant speed, accelerating and decelerating shocks (according
to Figure \ref{fig:xpow}). In addition, the behavior of the density
(either constant, vanishing or diverging) near the piston boundary
is evident (according to Figure \ref{fig:qpow}). \label{fig:RT}}
\end{figure*}

In Figure \ref{fig:RT} several $x-t$ diagrams are plotted for various
values of $\tau,\omega$. These diagrams describe the exact temporal
evolution of the locations of Lagrangian mass elements (via equation
\ref{eq:xmt}), through which the shock is propagating. One can see
that different values of $\tau,\omega$ result in constant speed,
accelerating and decelerating motion (according to Figure \ref{fig:xpow}).
In addition, it is seen that for some combinations of $\tau,\omega$,
the cells near the piston are compressed/expanded substantially, due
to the fact that the density near the piston boundary is either constant,
vanishing or diverging (according to Figure \ref{fig:qpow}).

\begin{figure*}[t]
\begin{centering}
\includegraphics[scale=0.45]{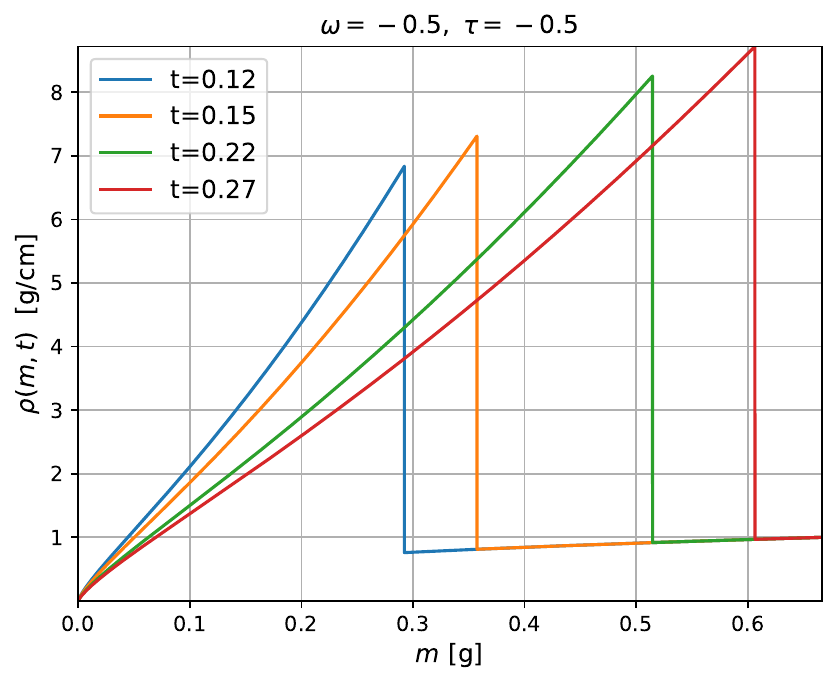}\includegraphics[scale=0.45]{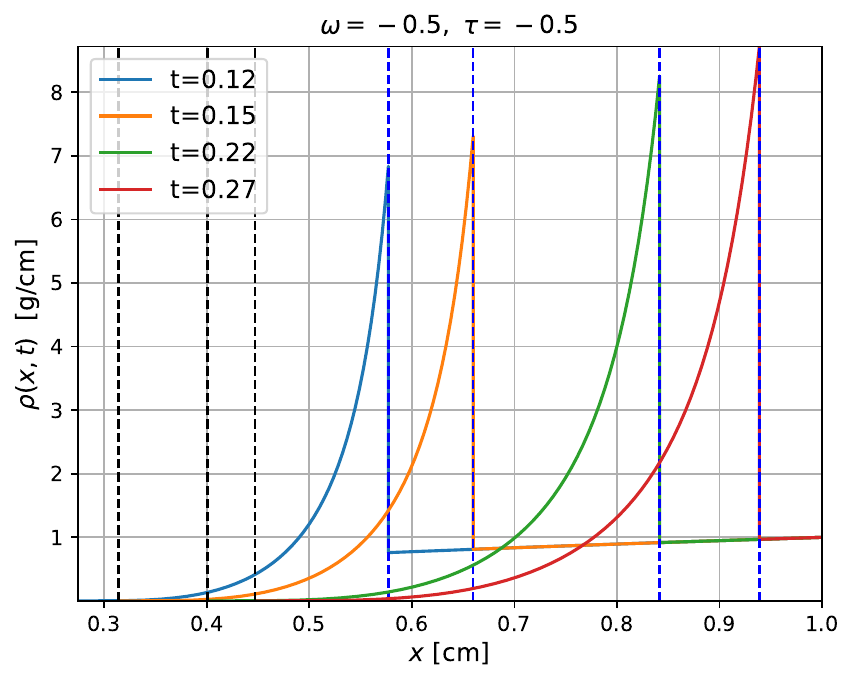}
\par\end{centering}
\begin{centering}
\includegraphics[scale=0.45]{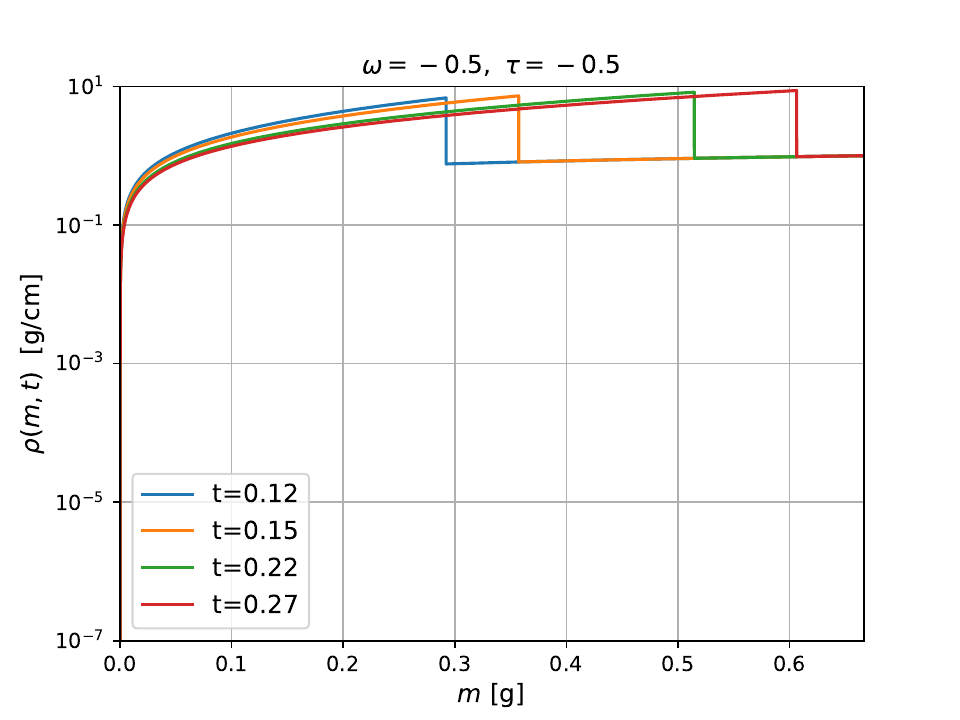}\includegraphics[scale=0.45]{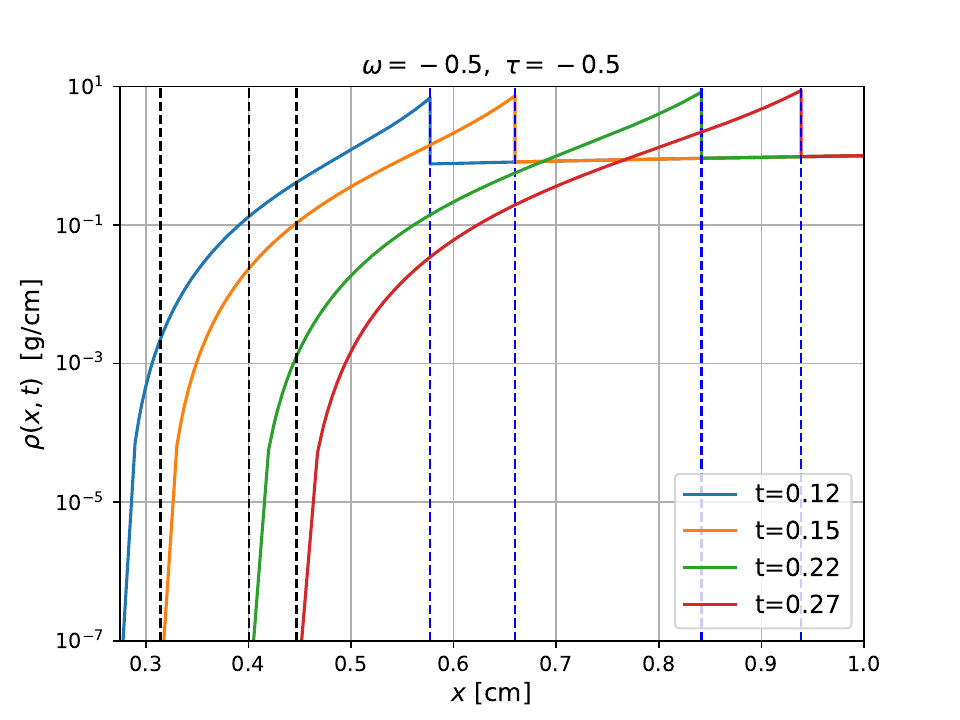}
\par\end{centering}
\begin{centering}
\includegraphics[scale=0.45]{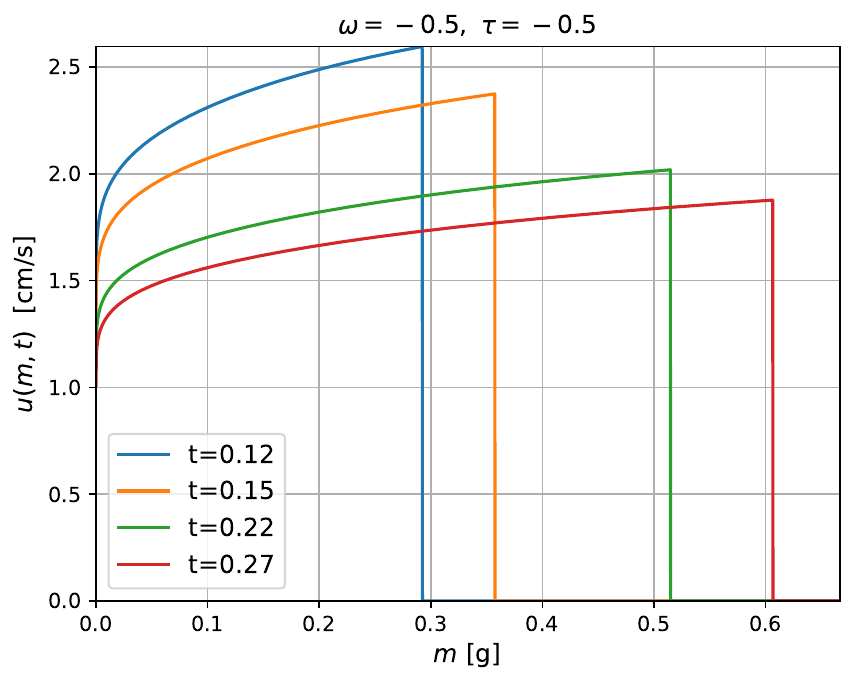}\includegraphics[scale=0.45]{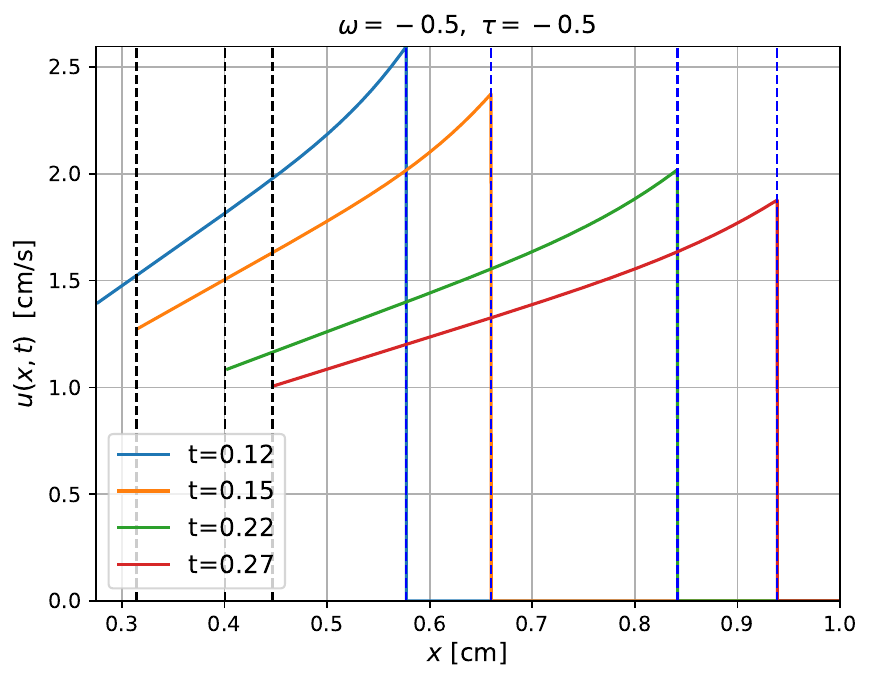}
\par\end{centering}
\begin{centering}
\includegraphics[scale=0.45]{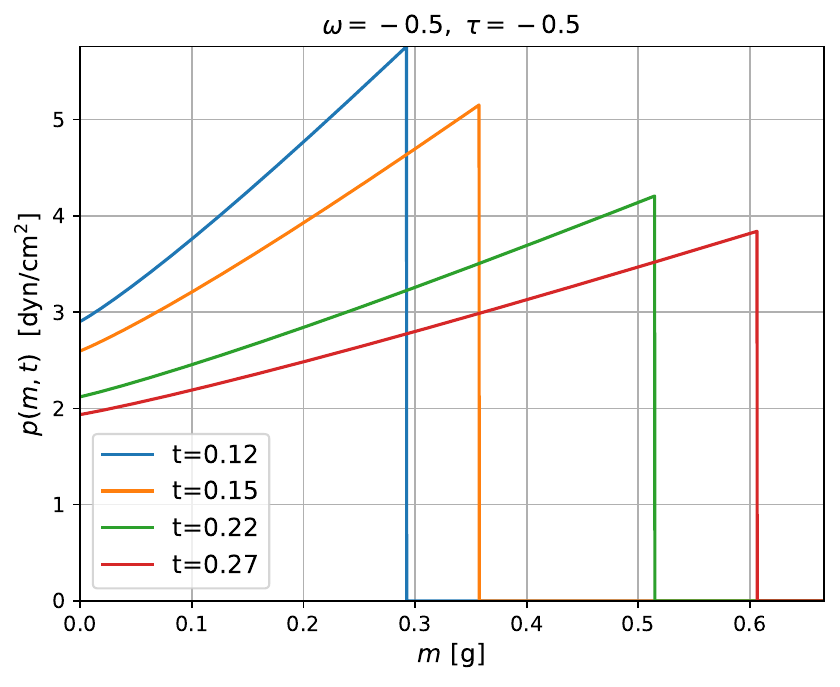}\includegraphics[scale=0.45]{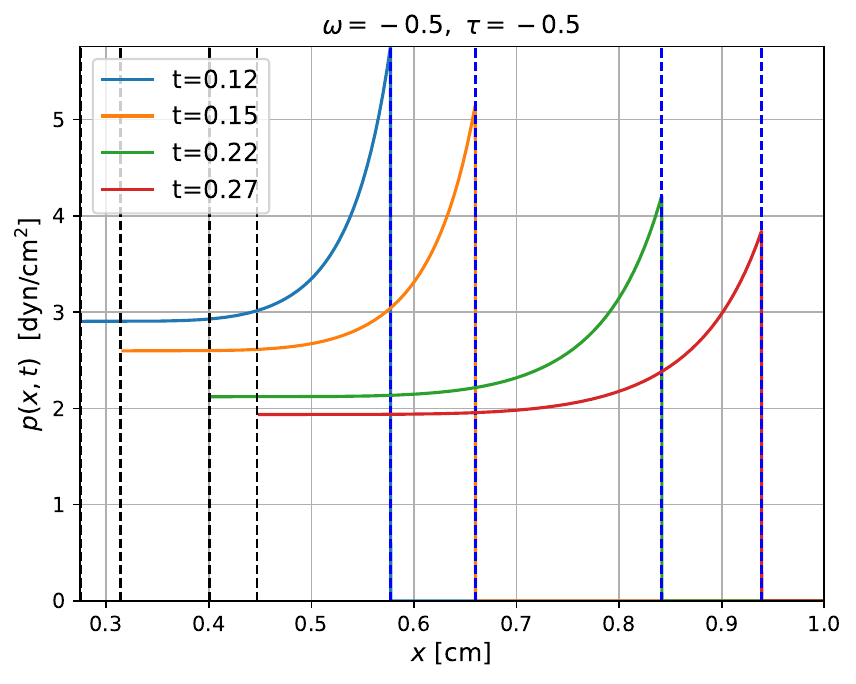}
\par\end{centering}
\caption{Hydrodynamic profiles for the density, pressure and velocity, at various
times (as written in the legends), for $\tau=-0.5$, $\omega=-0.5$,
$\gamma=1.25$, $\rho_{0}=1\text{g/cm}$ and $p_{0}=1\text{dyn/cm}^{2}\text{\textbf{/sec}}^{\tau}$.
The profiles are plotted with respect to Lagrangian mass coordinates
(left column) as well as in real space (right column), in which the
dashed black and blue lines represent the piston and shock positions,
respectively. In this case, according to Figure \ref{fig:qpow}, the
density vanishes near the piston, as can be seen in the density plots
which are also given in log scale.\label{fig:hyd_prof}}
\end{figure*}

\begin{figure*}[t]
\begin{centering}
\includegraphics[scale=0.45]{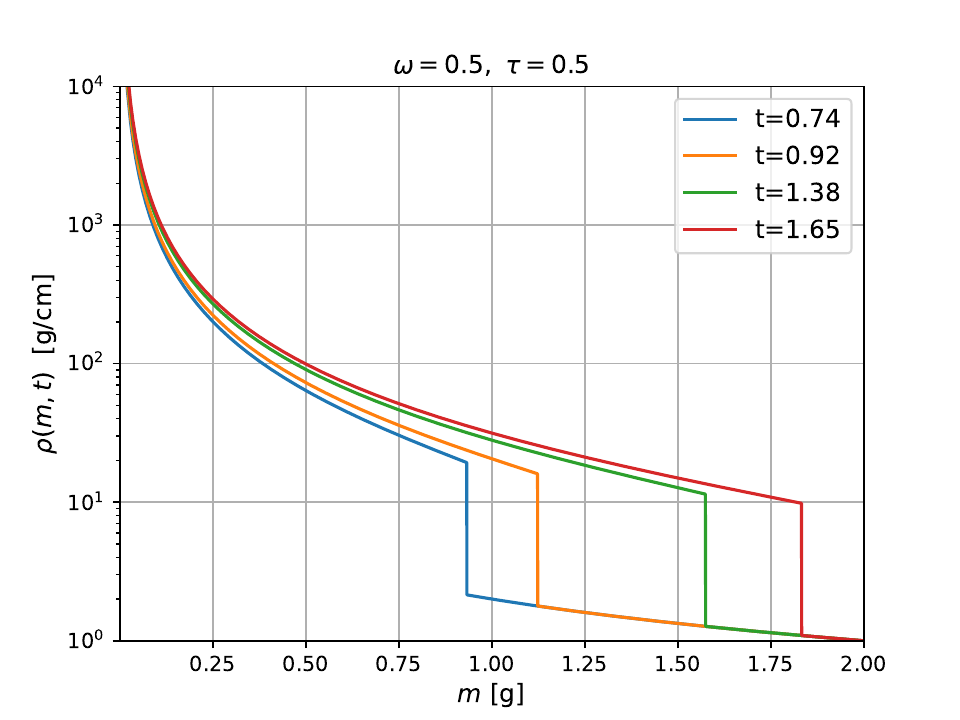}\includegraphics[scale=0.45]{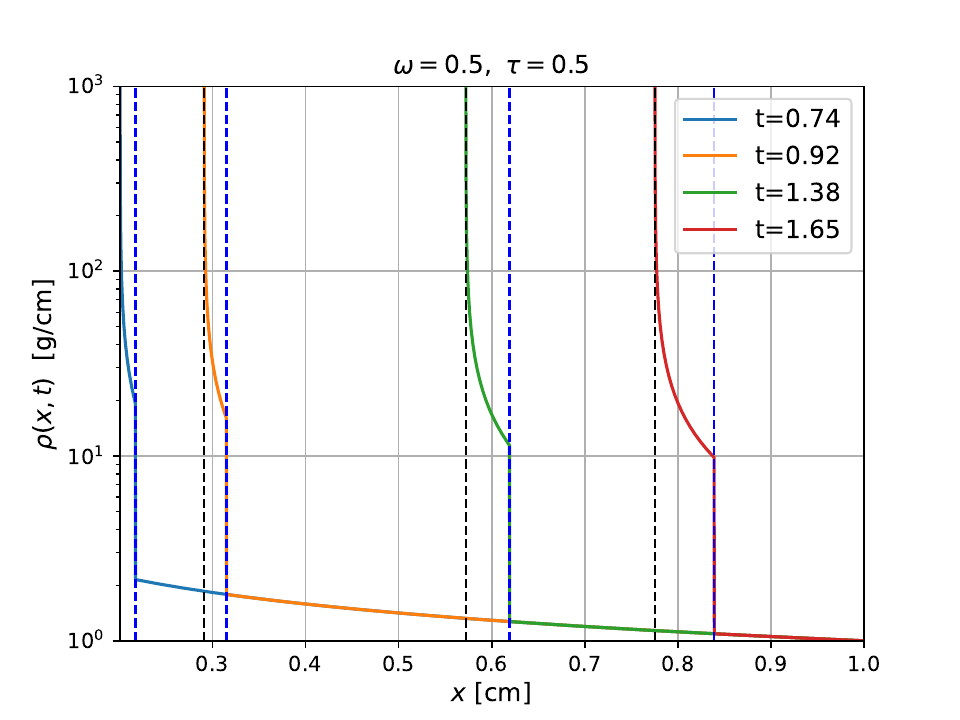}
\par\end{centering}
\begin{centering}
\includegraphics[scale=0.45]{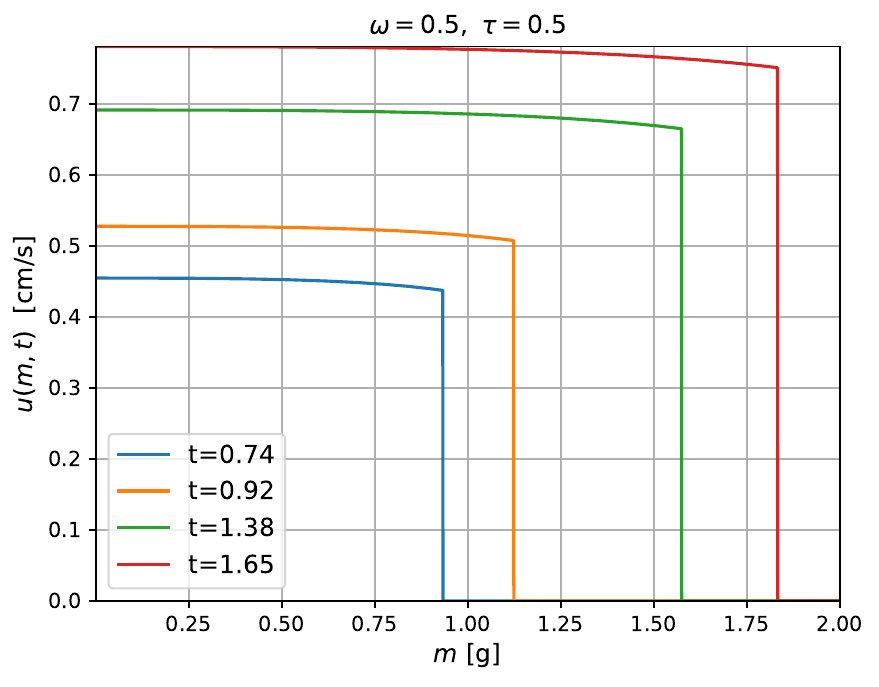}\includegraphics[scale=0.45]{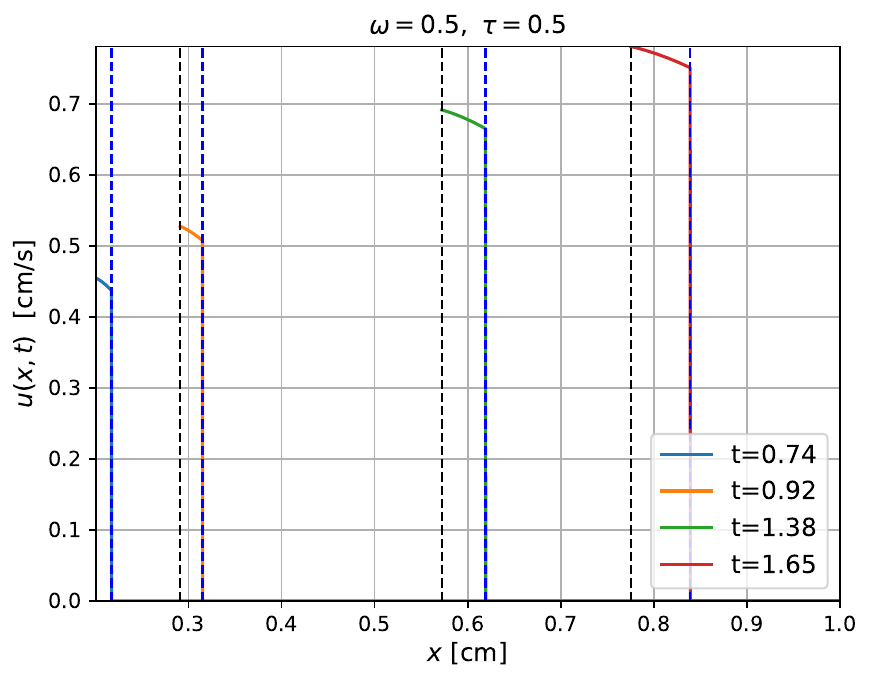}
\par\end{centering}
\begin{centering}
\includegraphics[scale=0.45]{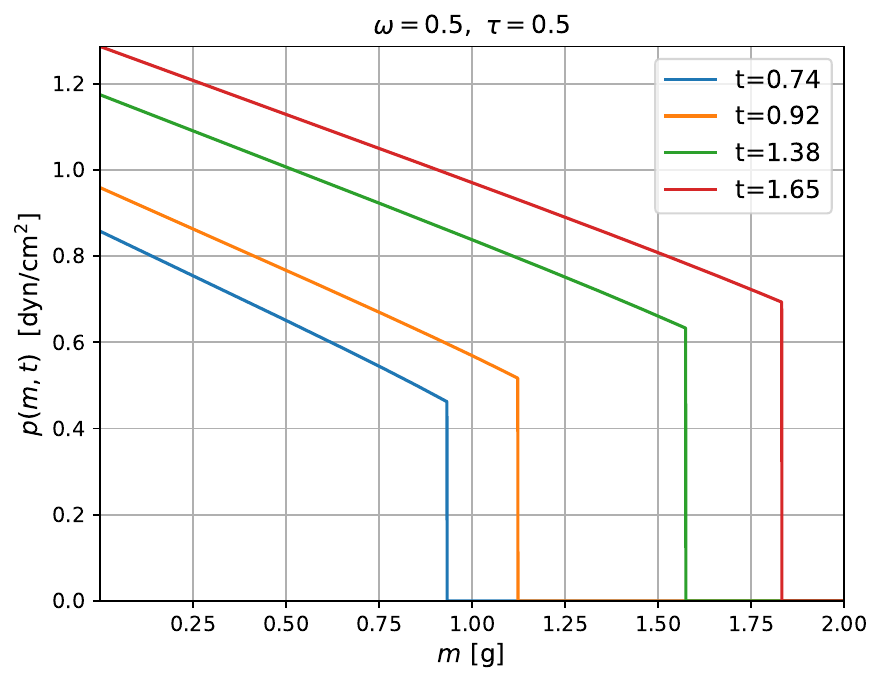}\includegraphics[scale=0.45]{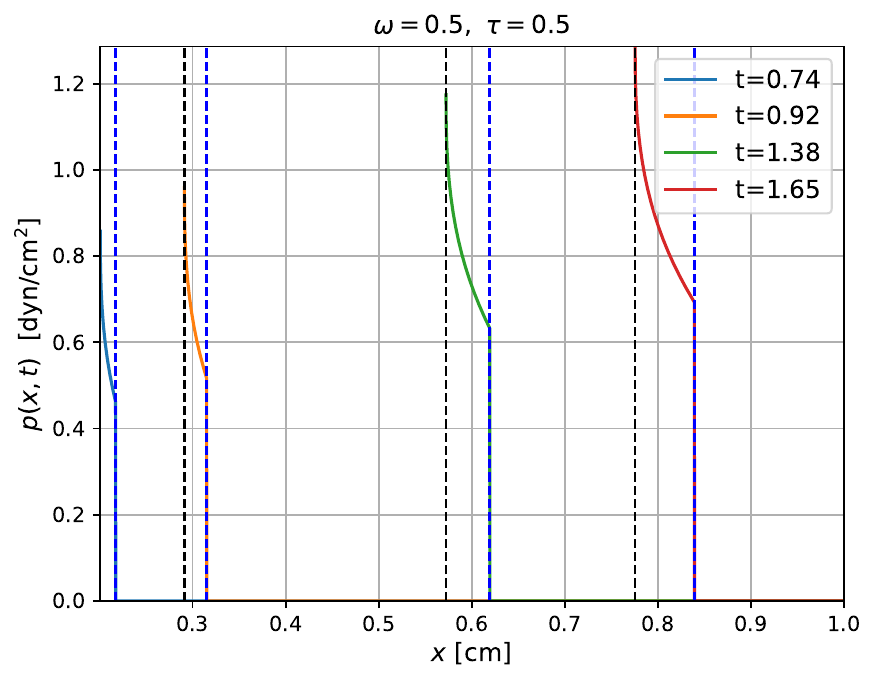}
\par\end{centering}
\caption{Same as Figure \ref{fig:hyd_prof}, but for $\tau=0.5$, $\omega=0.5$.
In this case, according to Figure \ref{fig:qpow}, the density diverges
near the piston, as can be seen in the density plots. \label{fig:hyd_prof1}}
\end{figure*}

In Figures \ref{fig:hyd_prof}-\ref{fig:hyd_prof1} we present the
evolution of hydrodynamics profiles in both Lagrangian in Eulerian
coordinates. The profiles were by obtained from the self-similar solution
(via equations \ref{eq:v_ss}-\ref{eq:p_ss}) for two cases: $\omega=\tau=-0.5$
and $\omega=\tau=0.5$. As can be seen from the Figures, in the first
case the density vanishes near the piston (according to Figure \ref{fig:qpow})
and the shock propagation slows down over time (also evident in Figure
\ref{fig:RT}), while for the latter case the density diverges near
the piston and the shock accelerates.



\section{The energy in the system}

\begin{figure}[t]
\begin{centering}
\includegraphics[scale=0.6]{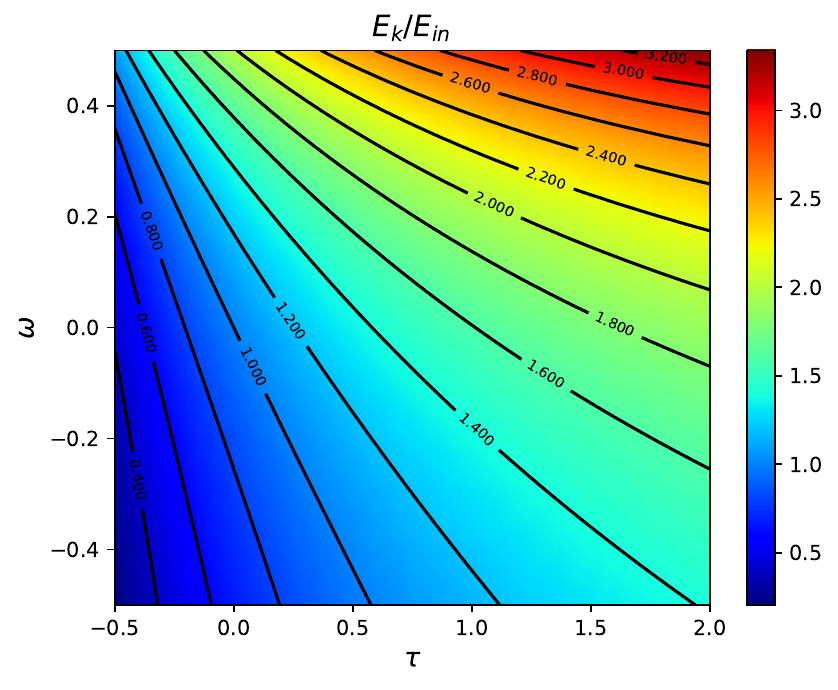}
\par\end{centering}
\caption{A color plot of the ratio between the total kinetic and internal energies
in the system (equation \ref{eq:erat}), as a function of $\tau,\omega$
for $\gamma=1.25$.\label{fig:erat}}
\end{figure}

The total internal and kinetic energy in the system are given by,
respectively:
\begin{equation}
E_{in}\left(t\right)=\frac{1}{r}\int_{0}^{m_{s}\left(t\right)}p\left(m,t\right)v\left(m,t\right)dm,
\end{equation}
\begin{equation}
E_{k}\left(t\right)=\int_{0}^{m_{s}\left(t\right)}\frac{1}{2}u^{2}\left(m,t\right)dm.
\end{equation}
By substituting the self-similar representation \ref{eq:v_ss}-\ref{eq:p_ss}
we find:

\begin{equation}
E_{in}\left(t\right)=v_{0}^{\frac{1}{2-\omega}}p_{0}^{\frac{3-\omega}{2-\omega}}t^{\frac{3\tau-\tau\omega+2}{2-\omega}}\int_{0}^{\xi_{s}}\frac{P\left(\xi\right)V\left(\xi\right)}{r}d\xi,\label{eq:ein}
\end{equation}

\begin{equation}
E_{k}\left(t\right)=v_{0}^{\frac{1}{2-\omega}}p_{0}^{\frac{3-\omega}{2-\omega}}t^{\frac{3\tau-\tau\omega+2}{2-\omega}}\int_{0}^{\xi_{s}}\frac{U^{2}\left(\xi\right)}{2}d\xi,\label{eq:ek}
\end{equation}
from which it is evident that the ratio between kinetic and internal
energy is time independent, and given by the numerical constant:
\begin{equation}
\frac{E_{k}}{E_{in}}=\frac{\int_{0}^{\xi_{s}}\frac{U^{2}\left(\xi\right)}{2}d\xi}{\int_{0}^{\xi_{s}}\frac{P\left(\xi\right)V\left(\xi\right)}{r}d\xi}.\label{eq:erat}
\end{equation}
This ratio is plotted as a function of $\tau,\omega$ in Figure \ref{fig:erat},
where the integrals were calculated numerically. The total energy
in the system is given by:

\begin{align}
E\left(t\right) & =E_{in}\left(t\right)+E_{k}\left(t\right)\\
= & v_{0}^{\frac{1}{2-\omega}}p_{0}^{\frac{3-\omega}{2-\omega}}t^{\frac{3\tau-\tau\omega+2}{2-\omega}}\int_{0}^{\xi_{s}}\left(\frac{P\left(\xi\right)V\left(\xi\right)}{r}+\frac{U^{2}\left(\xi\right)}{2}\right)d\xi.\nonumber 
\end{align}
We note that the integral expression can be calculated analytically,
in a manner similar the derivation of equation \ref{eq:vint}. To
that end, we use the similarity ODE for the conservation of total
energy (equation \ref{eq:etot}):
\begin{align}
\frac{PV}{r}+\frac{1}{2}U^{2} & =-\frac{1}{2}\left(\frac{2-\omega}{\omega+\tau}\right)\left(PU\right)'\\
 & -\frac{1}{2}\left(\frac{2+\tau}{\omega+\tau}\right)\left(\omega-1\right)\xi\left(\frac{PV}{r}+\frac{1}{2}U^{2}\right)'.\nonumber 
\end{align}
By a direct integration of this equation, we obtain an exact formula
for the total energy from the piston to a mass element $m$:
\begin{align}
 & \int_{0}^{\xi}\left(\frac{P\left(\xi'\right)V\left(\xi'\right)}{r}+\frac{1}{2}U^{2}\left(\xi'\right)\right)d\xi'\nonumber \\
= & \frac{\left(2+\tau\right)\left(1-\omega\right)}{3\tau-\tau\omega+2}\xi\left(\frac{P\left(\xi\right)V\left(\xi\right)}{r}+\frac{1}{2}U^{2}\left(\xi\right)\right)\nonumber \\
 & -\frac{2-\omega}{3\tau-\tau\omega+2}\left(P\left(\xi\right)U\left(\xi\right)-U\left(0\right)\right).
\end{align}
Using this identity, the total energy integral for the entire system
can be obtained by using the jump condition \ref{eq:hug_etot_ss}:
\begin{equation}
\int_{0}^{\xi_{s}}\left(\frac{P\left(\xi\right)V\left(\xi\right)}{r}+\frac{1}{2}U^{2}\left(\xi\right)\right)d\xi=\frac{2-\omega}{3\tau-\tau\omega+2}U\left(0\right).
\end{equation}
It is not surprising that $U\left(0\right)$ sets the total energy
in the problem, as it is the only quantity that detriments the piston
velocity (see equation \ref{eq:xpiston}).

\section{Comparison to Hydrodynamic Simulations\label{sec:Comparison-to-Hydrodynamic}}

\begin{figure*}[t]
\begin{centering}
\includegraphics[scale=0.5]{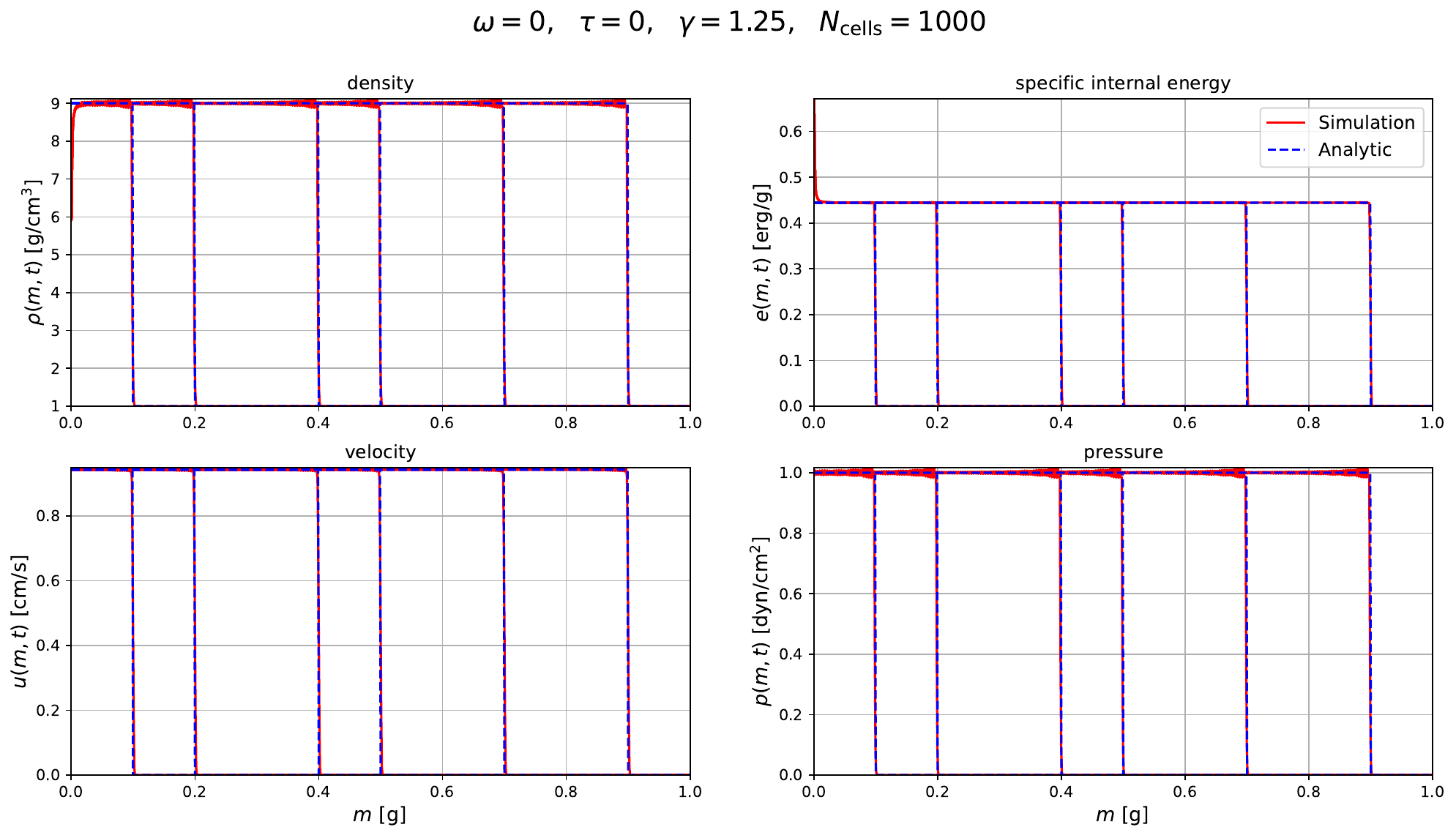}
\par\end{centering}
\caption{A comparison between the analytical solution (dashed blue lines) and
a numerical simulation (red lines) for the hydrodynamic profiles:
density (top left), specific internal energy (top right), velocity
(bottom left) and pressure (bottom right), as a function of the Lagrangian
mass coordinate $m$ for the parameters $\tau=0$, $\omega=0$, $\gamma=1.25$,
$\rho_{0}=1\text{g/cm}$ and $p_{0}=1\text{dyn/cm}^{2}/\text{sec}^{\tau}$.
The profiles are shown at various times where the shock positions
are at $0.1,0.2,0.4,0.5,0.7,0.9$ cm (as can be seen in Fig. \ref{fig:sim_euler_0_0}).
The simulation was performed with 1000 zones.\label{fig:sim_lag_0_0}}
\end{figure*}

\begin{figure*}[t]
\begin{centering}
\includegraphics[scale=0.5]{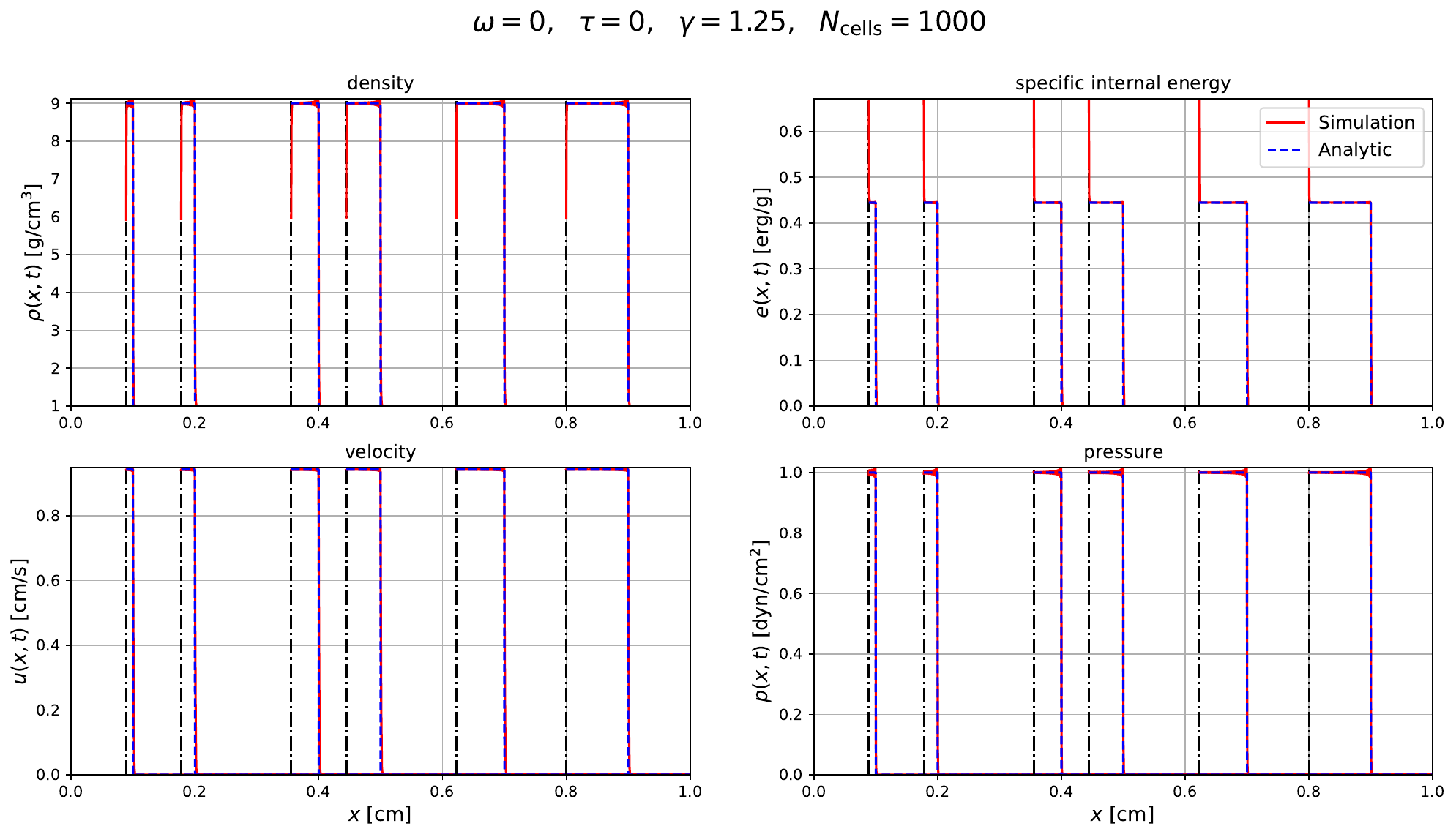}
\par\end{centering}
\caption{Same comparison as in Fig. \ref{fig:sim_lag_0_0}, with the hydrodynamic
profiles as a function of position (real space representation). The
piston positions are indicated by dashed-dotted vertical lines. \label{fig:sim_euler_0_0}}
\end{figure*}
\begin{figure*}[t]
\begin{centering}
\includegraphics[scale=0.5]{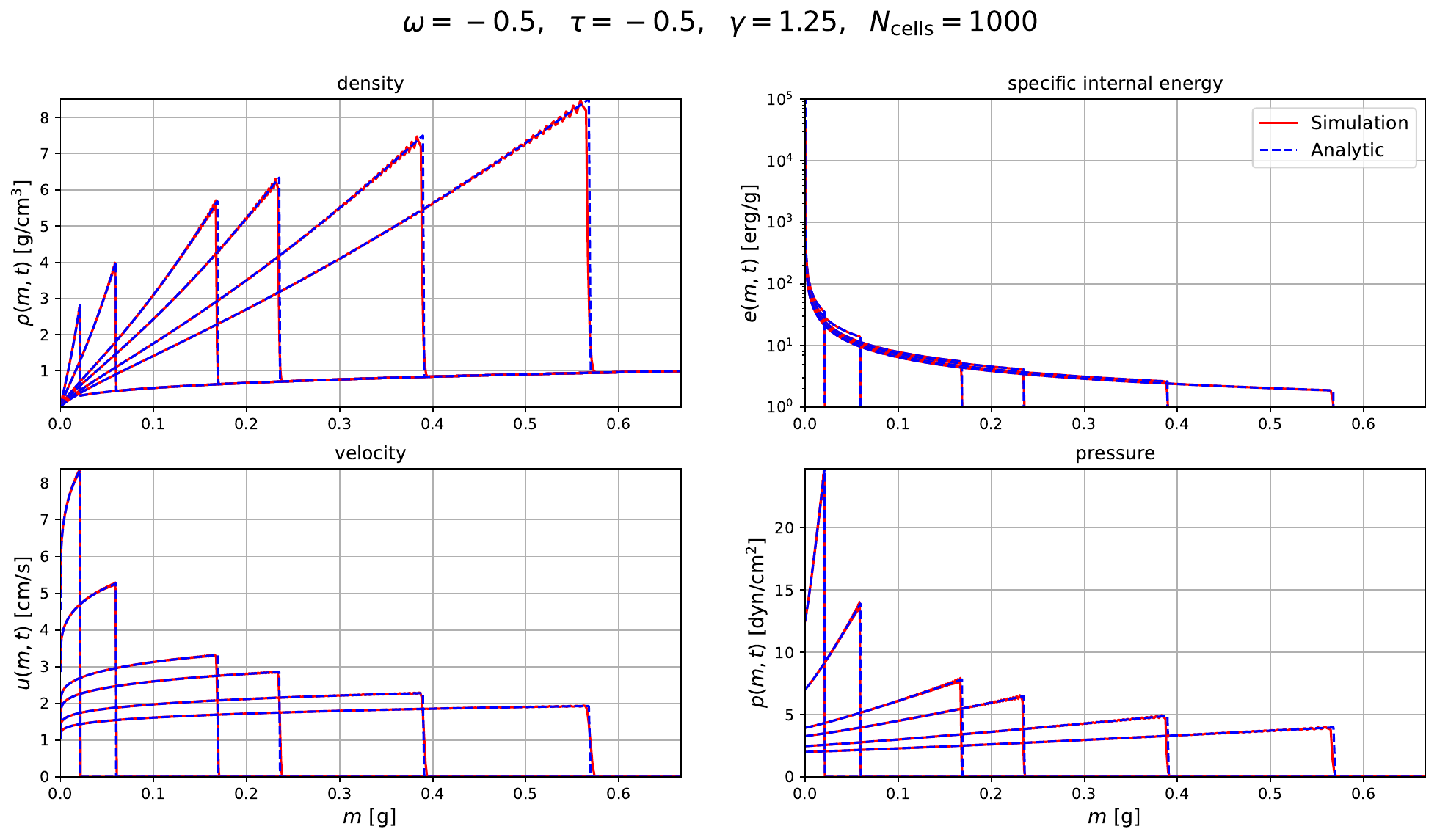}
\par\end{centering}
\caption{Same as Fig. \ref{fig:sim_lag_0_0}, for $\omega=-0.5$, $\tau=-0.5$.\label{fig:sim_lag_-0.5_-0.5}}
\end{figure*}
\begin{figure*}[t]
\begin{centering}
\includegraphics[scale=0.5]{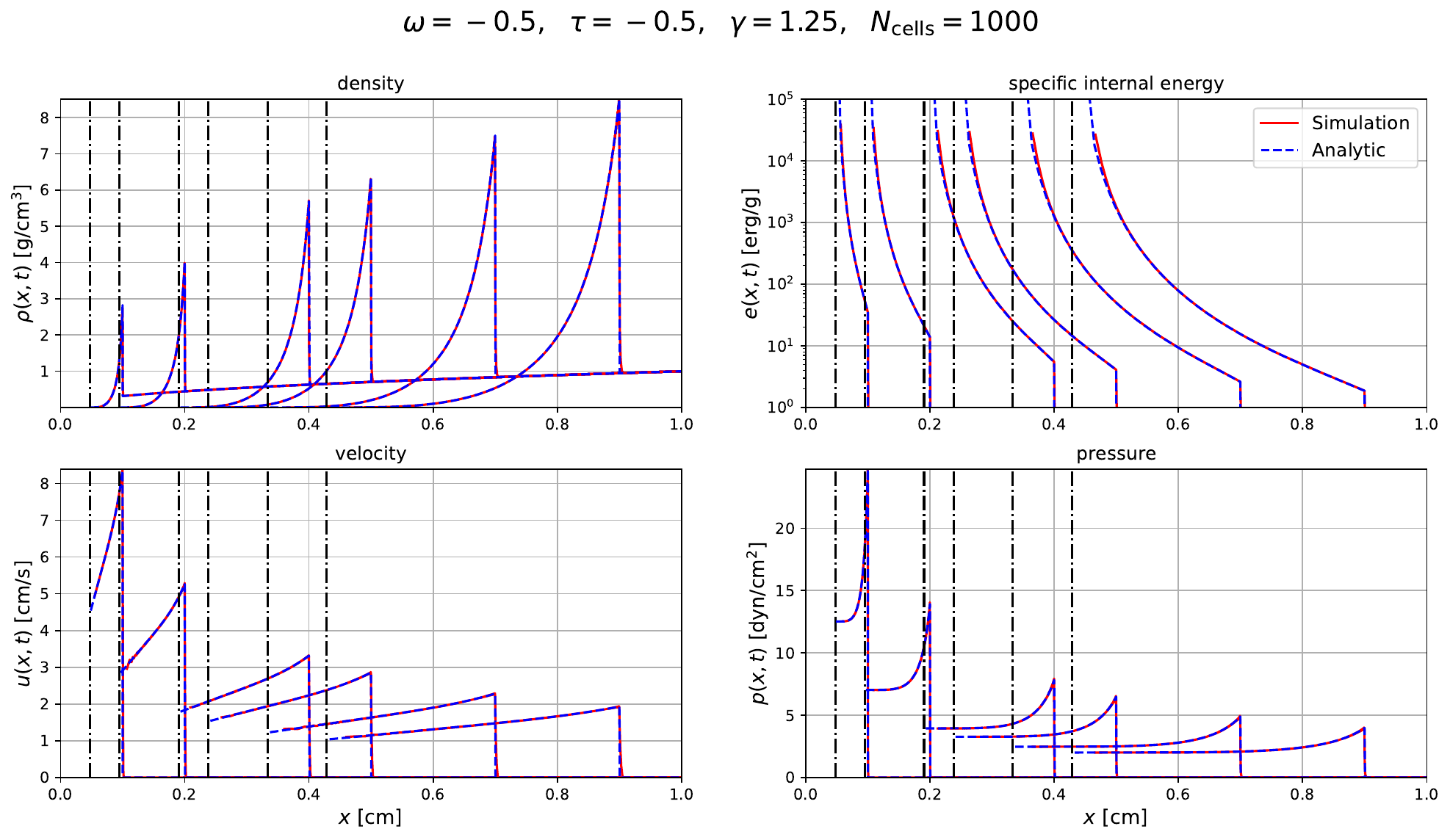}
\par\end{centering}
\caption{Same as Fig. \ref{fig:sim_euler_0_0}, for $\omega=-0.5$, $\tau=-0.5$.\label{fig:sim_euler_-0.5_-0.5}}
\end{figure*}

\begin{figure*}[t]
\begin{centering}
\includegraphics[scale=0.5]{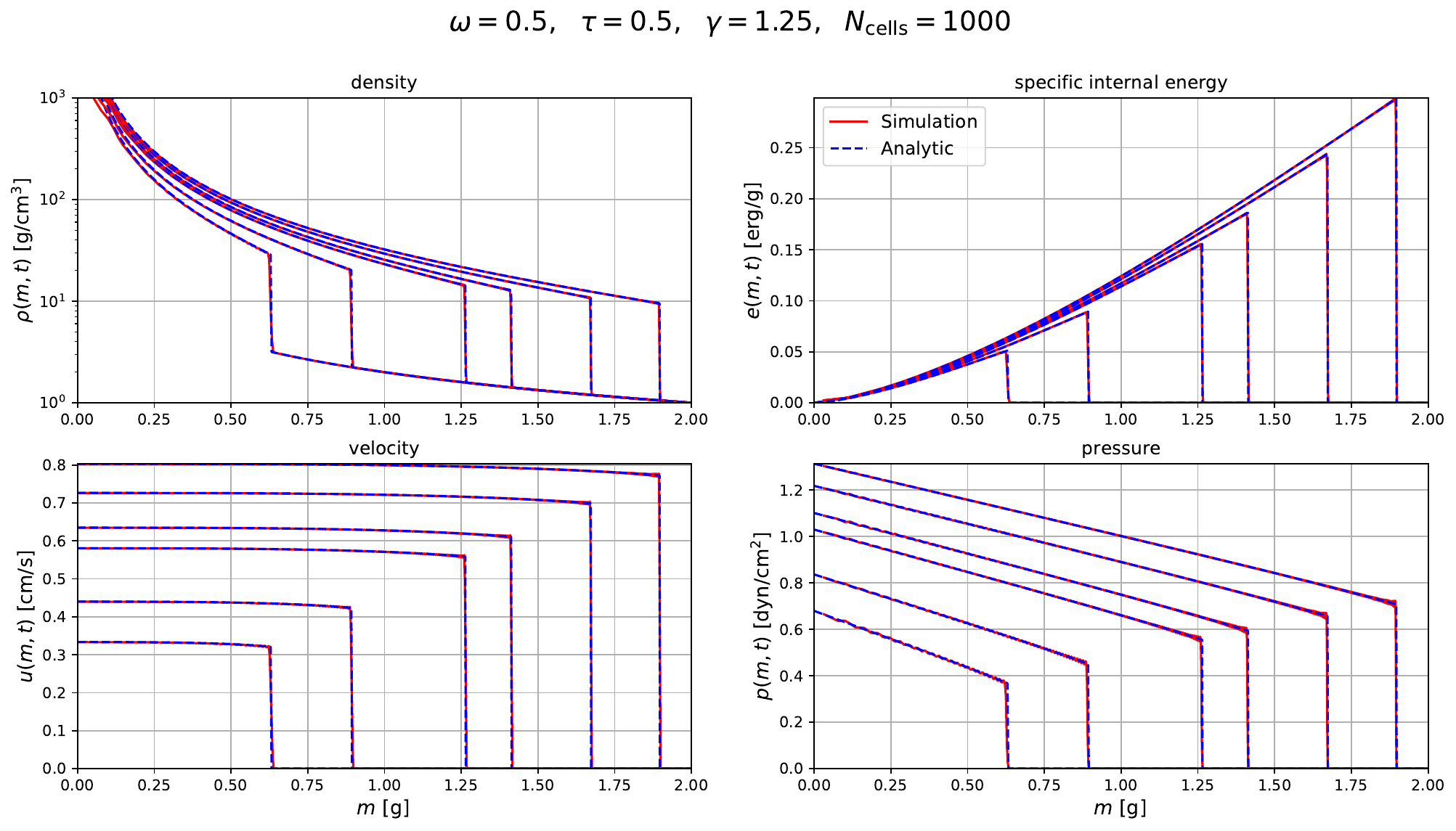}
\par\end{centering}
\caption{Same as Fig. \ref{fig:sim_lag_0_0}, for $\omega=0.5$, $\tau=0.5$.\label{fig:sim_lag_0.5_0.5}}
\end{figure*}
\begin{figure*}[t]
\begin{centering}
\includegraphics[scale=0.5]{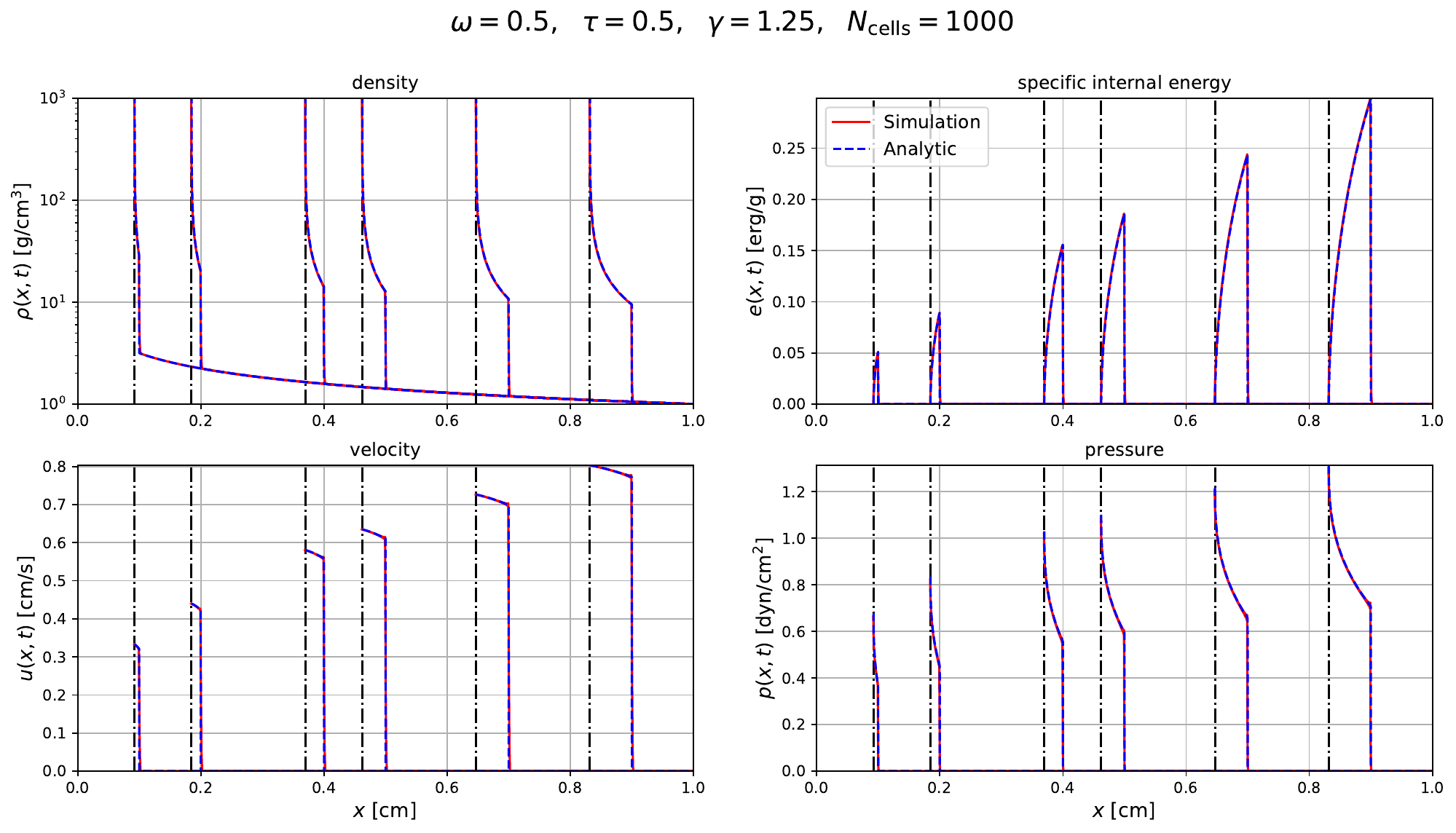}
\par\end{centering}
\caption{Same as Fig. \ref{fig:sim_euler_0_0}, for $\omega=0.5$, $\tau=0.5$.\label{fig:sim_euler_0.5_0.5}}
\end{figure*}

\begin{figure*}[t]
\begin{centering}
\includegraphics[scale=0.5]{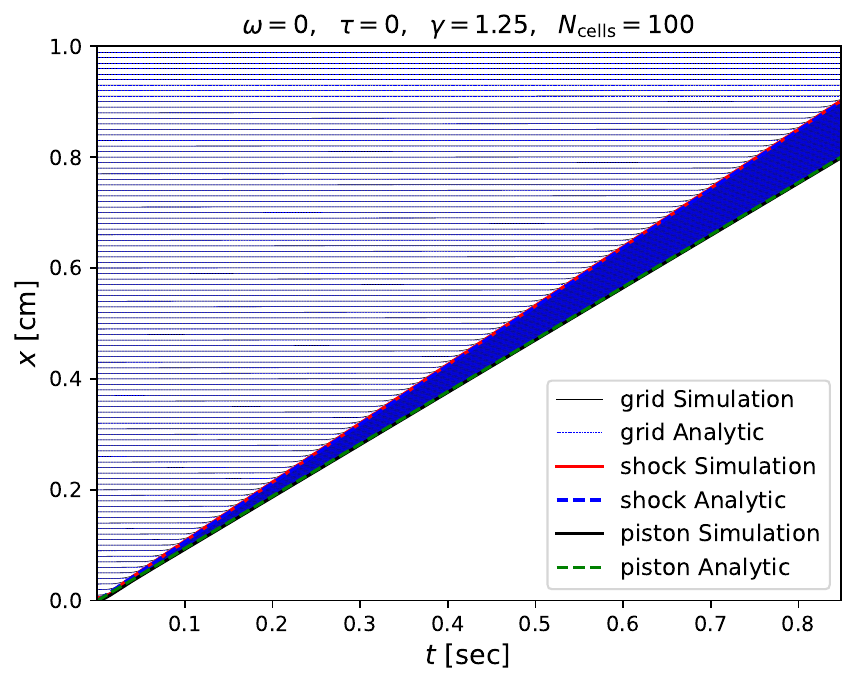}\includegraphics[scale=0.5]{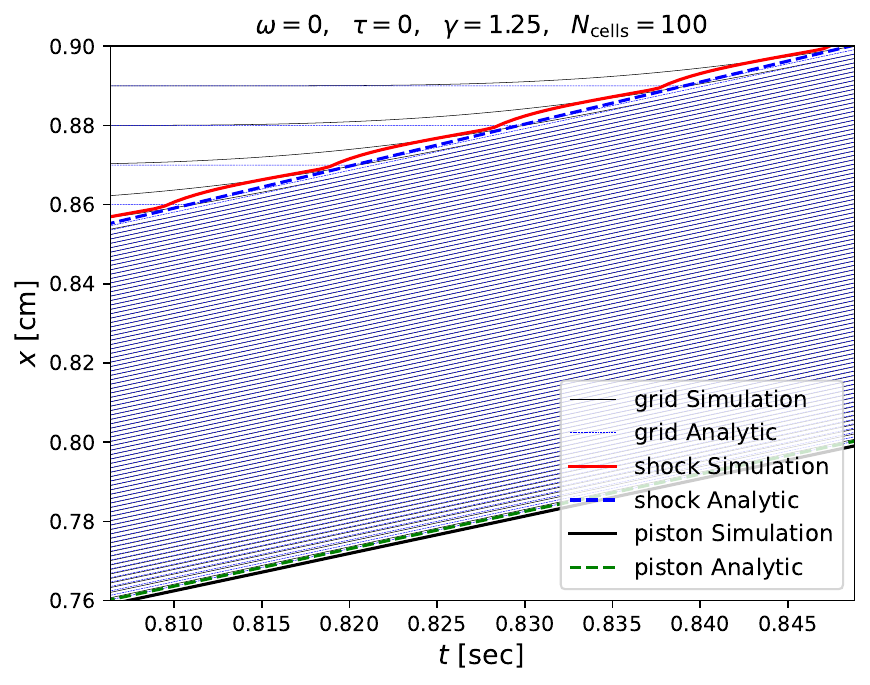}
\par\end{centering}
\begin{centering}
\includegraphics[scale=0.5]{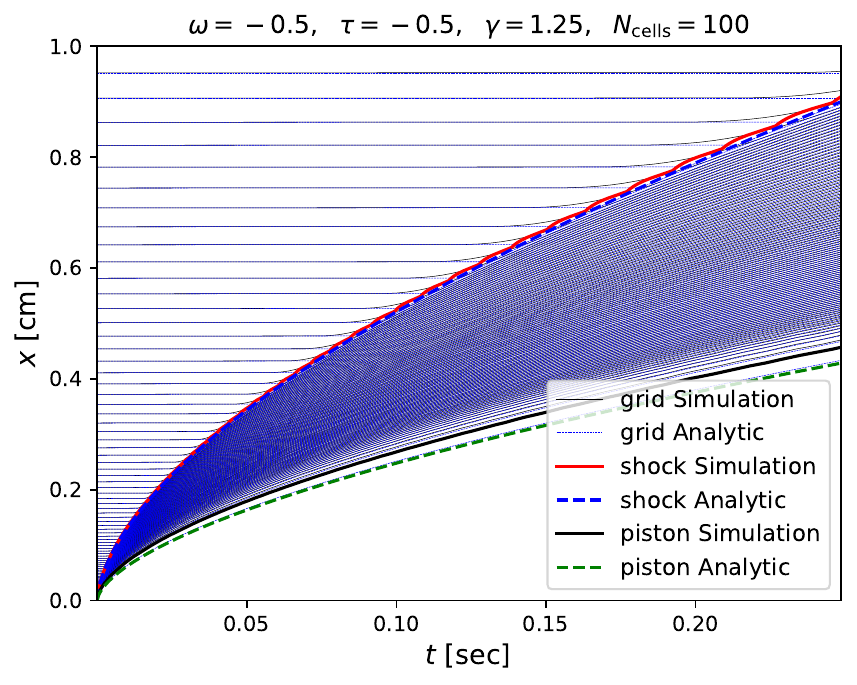}\includegraphics[scale=0.5]{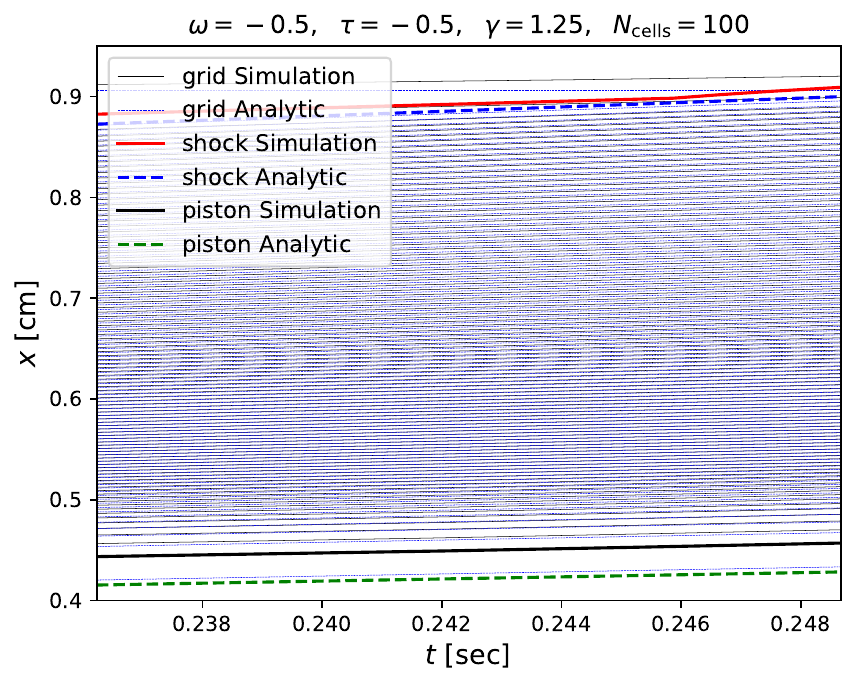}
\par\end{centering}
\begin{centering}
\includegraphics[scale=0.5]{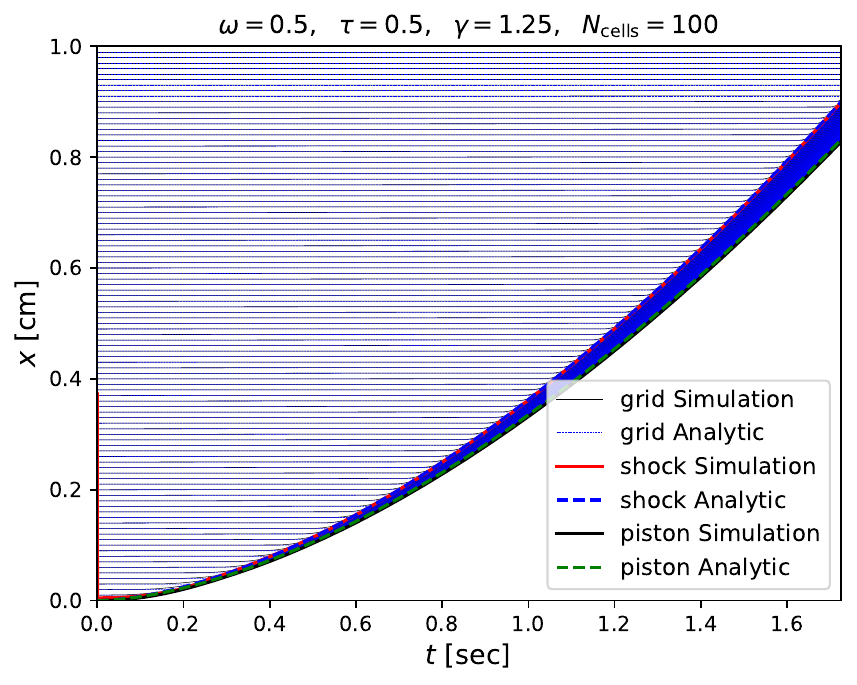}\includegraphics[scale=0.5]{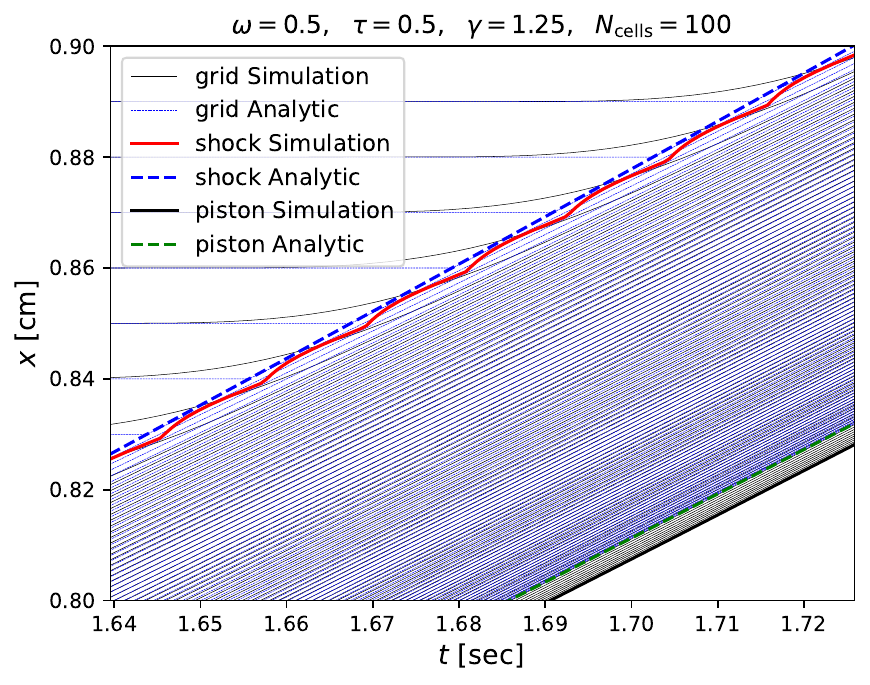}
\par\end{centering}
\caption{$x-t$ diagrams for the planar shock wave flow of 100 Lagrangian zones
for three cases: $\tau=0$, $\omega=0$ (upper figures), $\tau=-0.5$,
$\omega=-0.5$ (middle figures) and $\tau=0.5$, $\omega=0.5$ (lower
figures), where in all cases $\gamma=1.25$, $\rho_{0}=1\text{g/cm}$
and $p_{0}=1\text{dyn/cm}^{2}/\text{sec}^{\tau}$. We compare the
analytical solutions given by Eq. \ref{eq:xmt} (thin blue lines)
with numerical simulations (thin black lines). The location of the
shock front (analytic - dashed blue line, simulation - red thick line)
and the piston (analytic - dashed green line, simulation - black thick
line), are also shown. The figures on the right give a closer look
on the shocked region. \label{fig:sim_rt}}
\end{figure*}
\begin{figure*}[t]
\begin{centering}
\includegraphics[scale=0.25]{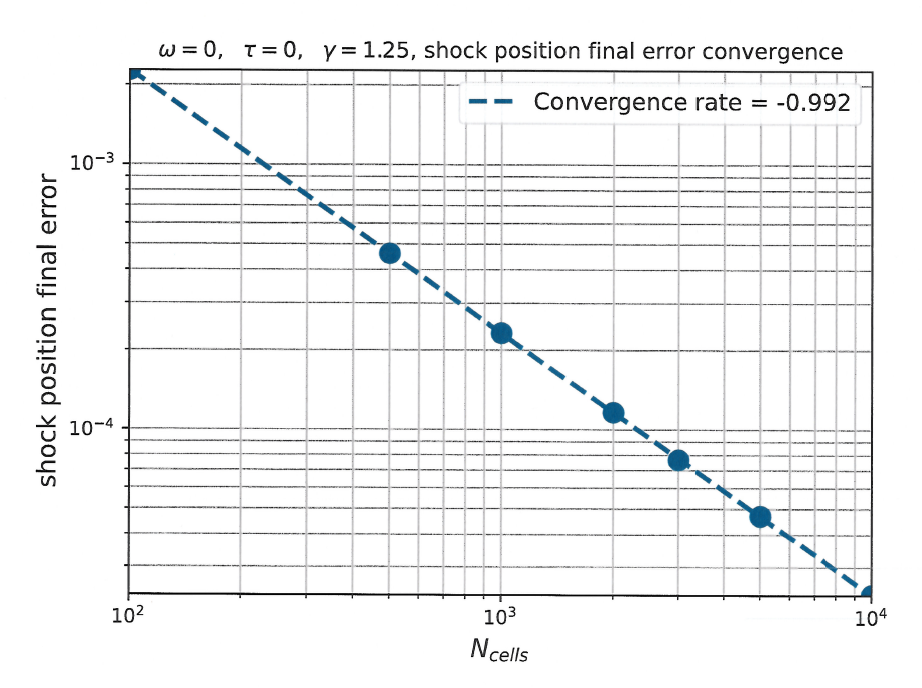}\includegraphics[scale=0.25]{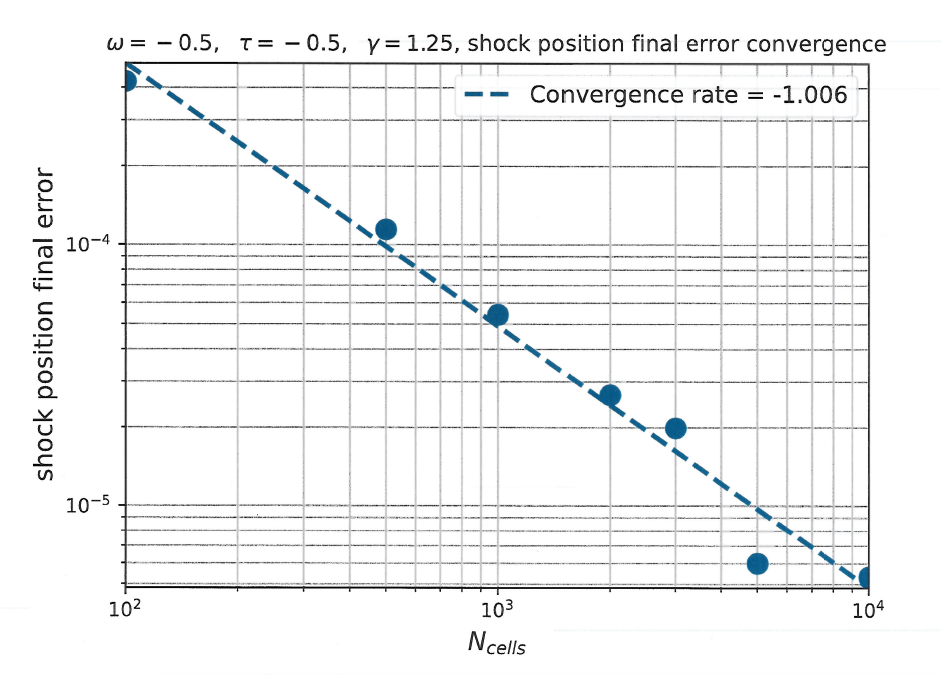}\includegraphics[scale=0.25]{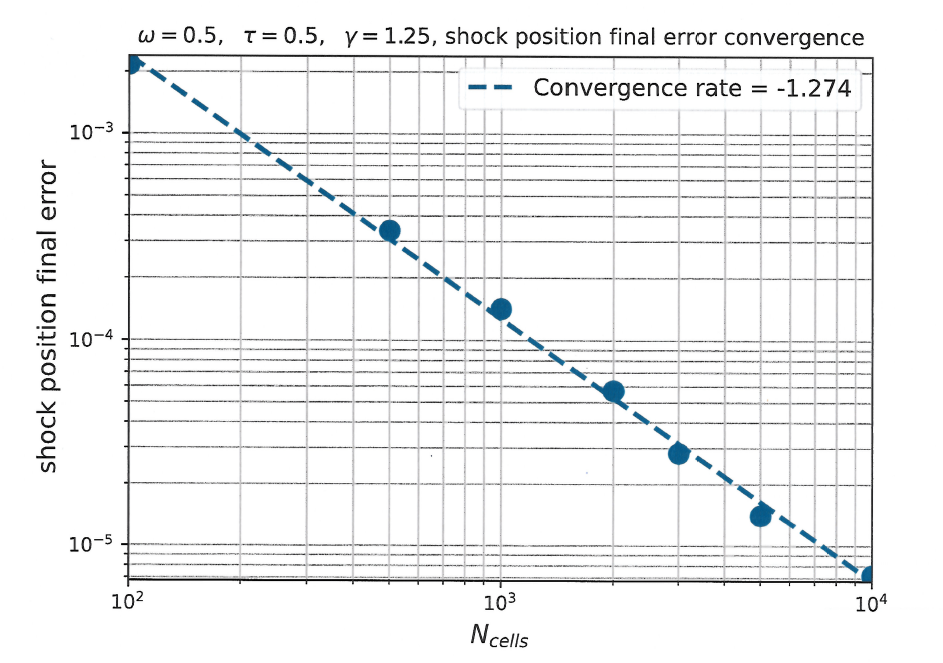}
\par\end{centering}
\caption{Spatial convergence plots for the shock position at the end of the
three simulations defined in section \ref{sec:Comparison-to-Hydrodynamic}.
The values of $\omega,\tau$ are listed in the titles. Shown are the
relative error between the final shock positions as obtained in the
simulations and the analytic result, as a function of the number of
computational zones $N_{cells}$. The (fitted) convergence rate, which
is given by the slope of the convergence line in log-log scale, is
written in the legends of each figure. \label{fig:conv}}
\end{figure*}
In this section we compare the semi-analytical planar shock solutions
to numerical hydrodynamic simulations. We employ a standard staggered-mesh
artificial-viscosity one-dimensional fully Lagrangian code (which
is described in detail in appendix D of Ref. \cite{giron2021solutions}).
The pressure piston boundary condition is applied by setting a ghost
cell adjacent to the leftmost zone with a temporally applied pressure
$p_{\text{ghost}}\left(t\right)=p_{0}t^{\tau}$. The initial pressure
and velocity are zero throughout the system, while the initial density
is set according to the spatial power-law \ref{eq:rho_init}. 

We compare the analytical solutions and the simulated results for
the density, velocity, pressure and specific internal energy profiles,
which are represented in mass coordinates as well as in real space.
$1000$ computational zones are used in these simulations. In addition,
in Fig. \ref{fig:sim_rt} we compare the analytical $x-t$ diagrams
(see section \ref{sec:x-t}), to simulations with a lower number of
$100$ computational zones.

All simulation are performed with the parameters $\gamma=1.25$, $\rho_{0}=1\text{g/cm}$
and $p_{0}=1\text{dyn/cm}^{2}/\text{sec}^{\tau}$. The final simulation
time is defined by the time for which the analytical shock position
is at $0.9\text{cm}$ (see eq. \ref{eq:xshock}).\textbf{ }We define
3 test cases. 
\begin{enumerate}
\item $\tau=0$, $\omega=0$ - this is the trivial case of a constant applied
pressure, which generates a shock which propagates into a homogeneous
cold medium. All hydrodynamic profiles are constant in space (via
the trivial shock jump conditions \ref{eq:vs_hug}-\ref{eq:ps_hug}),
as shown Figures \ref{fig:sim_lag_-0.5_-0.5}-\ref{fig:sim_euler_-0.5_-0.5}.
The shock velocity is constant in time (see equation \ref{eq:xshock}).
The total energy increases linearly in time, and the internal and
kinetic energy are equal (see equations \ref{eq:ein}-\ref{eq:erat}
and Figure \ref{fig:erat}).
\item $\tau=-0.5$, $\omega=-0.5$ - in this case the applied pressure decreases
over time, and the generated shock propagates into a cold medium with
a decreasing density profile. According to Figure \ref{fig:qpow},
the density vanishes near the piston, as shown in Figures \ref{fig:sim_lag_-0.5_-0.5}-\ref{fig:sim_euler_-0.5_-0.5}.
According to equation \ref{eq:xshock} (see also Figure \ref{fig:xpow}),
the shock propagation slows down over time as $x_{s}\left(t\right)\propto t^{0.6}$
(also evident in Figure \ref{fig:sim_rt}). In order to properly resolve
the diminishing density near the piston, in this case the simulations
were performed with non uniform (geometrically) spaced zones where
the initial zone widths were set according to $\Delta x_{n}=q\Delta x_{n-1}$
with $q=1.002$ for the fine simulations (with 1000 zones), and $q=1.05$
for the coarser simulations (with 100 zones). This geometrically spaced
grid is evident in the $x-t$ plots in Figure \ref{fig:sim_rt}.
\item $\tau=0.5$, $\omega=0.5$ - in this case the applied pressure increases
over time, and the generated shock propagates into a cold medium with
an increasing density profile. According to Figure \ref{fig:qpow},
the density diverges near the piston, as shown in Figures \ref{fig:sim_lag_0.5_0.5}-\ref{fig:sim_euler_0.5_0.5}.
According to equation \ref{eq:xshock} (see also Figure \ref{fig:xpow}),
the shock propagation accelerates over time as $x_{s}\left(t\right)\propto t^{5/2}$
(also evident in Figure \ref{fig:sim_rt}). 
\end{enumerate}
As can be seen in Figures \ref{fig:sim_lag_0_0}-\ref{fig:sim_rt},
in all cases we observe a very good agreement between the analytical
solutions and the numerical simulations.\textbf{ }In Figure \ref{fig:conv},
we show the spatial convergence of the numerical simulations to the
analytic results for the final shock position (which is $0.9$cm by
construction). The resulting order of convergence, which is obtained
by the slope of the relative error in log-log scale, is about $1.0$
(that is, a linear convergence) for the cases $\omega=0,\ \tau=0$
and $\omega=-0.5,\ \tau=-0.5$, and about 1.3 (slightly better than
linear) for the case $\omega=0.5,\ \tau=0.5$. 

\section{Summary}

In this work we have generalized the solutions to the compressible
hydrodynamics problem of piston driven shock waves in homogeneous
planar ideal gas media, to non-homogeneous media with an initial density
profile in the from of a spatial power law. Similarity solutions were
developed systematically in both Lagrangian and Eulerian coordinates.
It was shown that the generalized solutions take various qualitatively
different forms according to the values the temporal piston pressure
exponent $\tau$ and the spatial initial density exponent $\omega$.
Specifically, it was shown that there exist different families of
solutions for which: (i) the shock propagates at constant speed, accelerates
or slows down (ii) the density near the piston is either finite, vanishes
or diverges and (iii) the total kinetic energy can be larger, equal
or smaller than the total internal energy. 

An elaborate comparison between the semi-analytical planar shock solutions
to numerical hydrodynamic simulations was performed, and a very good
agreement was reached. This highlights the use of the new solutions
for the purpose of verification and validation of numerical hydrodynamics
simulation codes.

Finally, we note that the solutions presented in this work, can be
used directly to generalize the ablation driven shock wave solutions
\cite{shussman2015full,heizler2021radiation}, to media which have
a non-homogeneous initial density profile. 

\subsection*{Availability of data}

The data that support the findings of this study are available from
the corresponding author upon reasonable request.

\bibliographystyle{unsrt}
\bibliography{datab}

\pagebreak
\end{document}